\documentclass[onecolumn,a4paper]{autart}    

\usepackage{graphicx}          
\usepackage[dvips]{epsfig}    

\usepackage{mathptmx} 
\usepackage{times} 

\usepackage{amsmath} 
\usepackage{amssymb}  
\graphicspath{{Images/}}

\usepackage{color}
\usepackage{wrapfig}

\usepackage{cutwin}
\usepackage{soul}

\usepackage{listings}                             
\definecolor{mygreen}{RGB}{28,172,0} 
\definecolor{mylilas}{RGB}{170,55,241}

\newtheorem{Lemma}{Lemma}

\newtheorem{assumption}{Assumption}  
\newtheorem{Theorem}{Theorem}
\newtheorem{corollary}{Corollary}

\begin{document}

\lstset{language=Matlab,%
    breaklines=true,%
    morekeywords={matlab2tikz},
    keywordstyle=\color{blue},%
    morekeywords=[2]{1}, keywordstyle=[2]{\color{black}},
    identifierstyle=\color{black},%
    stringstyle=\color{mylilas},
    commentstyle=\color{mygreen},%
    showstringspaces=false,
    numbers=left,%
    numberstyle={\tiny \color{black}},
    numbersep=9pt, 
    emph=[1]{for,end,break},emphstyle=[1]\color{red}, 
}

\begin{frontmatter}

\title{
Cohesive 
Networks using 
Delayed Self Reinforcement 
}

\author{Santosh Devasia}\ead{devasia@uw.edu}    

\address{
Mechanical Engineering Department, 
U. of Washington, Seattle, USA 98195-2600}  


\begin{abstract}                          
How a network gets to the goal (a consensus value) can 
be as important as reaching the consensus value. 
While prior  methods focus on rapidly getting to a new consensus value,  
maintaining cohesion, during the transition between consensus values or during tracking, 
remains challenging and has not been addressed. 
The main contributions of this work are to address the problem of maintaining cohesion by: (i) proposing a new delayed self-reinforcement (DSR) approach; (ii) extending it 
for use with agents that have  higher-order, heterogeneous dynamics,
and (iii) developing stability conditions for the DSR-based method. 
With DSR, each agent uses current and past information from neighbors to infer the overall goal, and modifies the update law to improve cohesion.
The advantages of the proposed DSR approach are that it only requires  already-available information from a given network to improve the cohesion,  and does not require network-connectivity modifications (which might not be always feasible) nor 
increases in the system's overall response speed (which can require larger input). 
Moreover,  illustrative simulation examples are used to comparatively evaluate the performance with and without DSR. 
The simulation results show substantial  improvement in cohesion with DSR.

\end{abstract}
\end{frontmatter}

\section{Introduction}
  \vspace*{-0.1in}\noindent
The goal of this work is to enable cohesive transitions  in multi-agent networks, e.g., to enable similar response in each agent when 
transitioning from one consensus value to another. 
How a network gets to the final goal (i.e., cohesion during the transition) can be as important as reaching the 
final goal, e.g., to maintain specified inter-vehicle spacing in  connected, automated transportation systems~\cite{Ren_07,TALEBPOUR2016}, 
to align the orientation of agents  during maneuvers of  flocks and swarms in nature, e.g.,~\cite{Huth_92,Vicsek_95,Attanasi_14}, 
and to maintain formation of engineered networks such as satellites, unmanned autonomous vehicles and  
collaborative robots~\cite{Jadbabaie_03,Ren_Beard_05,Olfati_Saber_06,Wang_13,Zhou_satellite,Wang_containment_2018}. 
While prior  methods aim to achieve rapid convergence to a consensus goal, 
cohesion during transitions between the consensus values has not been addressed and remains challenging. 
The loss of cohesion during transitions arises  because information about the desired change 
(such as the desired orientation or speed)  in the goal 
might be available to only a few agents in the network
and needs to diffuse through the system.  
The resulting response-time delays, between agents that are \lq\lq{close}\rq\rq\  to the information source in the network and those that are \lq\lq{farther away}\rq\rq,  lead to loss of   cohesion during the transition, even though they all reach the final  goal.  
The impact of transition-cohesion loss   
can be mitigated by using additional control effort,  e.g., designed 
to maintain inter-agent spacing and reduce formation distortions~\cite{Jadbabaie_03,Ren_Beard_05,Olfati_Saber_06}. 
Nevertheless, a cohesive  transient response, when feasible, 
reduces the need for such additional control effort. Moreover, in biological systems,  the alignment response is transmitted faster 
than neighbor-to-neighbor  re-arrangements~\cite{Attanasi_14}, which indicates that cohesion improvements (e.g., in the alignment response) 
during the transitions might be more effective than 
the slower rearrangements of the agents (or other actions)  to correct for the  loss of cohesion during rapid maneuvers. 
This potential reduction in overall effort by maintaining cohesion motivates the current study aimed at improving transition cohesion.

Faster convergence to the new consensus value can improve transition cohesion. 
For example, faster convergence implies a smaller  
overall settling time of the network response during transitions between consensus values. Here, the settling time is the time needed for all agents 
to reach within a specified percentage of a final consensus state $Z=Z_f$ when transitioning from an initial consensus state $Z=Z_i$, where all agents have the same initial value. 
Faster settling reduces the potential delays between the responses of the agents, and in this sense, promotes cohesion.  
For example, the network's response can be speeded up when the network dynamics has the form, 
\begin{align}
 \dot{Z} (t)    = U(t) =  - \gamma \hat{K}Z (t) +\gamma \hat{B}  z_s (t) 
 \label{intro_sys_eq}
\end{align}
by scaling up the gain $\gamma$ of the network interactions, where $K = \gamma \hat{K}$ is the graph Laplacian and $z_s$ is the desired response. Nonlinear methods have also been proposed to achieve linear and finite time 
convergence, e.g.,~\cite{CORTES20061993,Bu_2018,Shihua_11,Hu_2019}. 
However, in general, increasing the overall speed of the network requires larger inputs $U$. 
Therefore, maximum-input constraints on the actuators can lead to restrictions on the  maximum response-speed increase, which in turn limits the 
achievable cohesion.

The response speed, and therefore, transition cohesion can be improved if there is choice in the structure of the network. 
For example, if  spatially-distant agents can be connected, 
then the information about changes in the desired response $z_s$ can spread faster, which can improve cohesion of agent responses. 
Similarly, a faster response can be achieved by optimally selecting the Laplacian  $ K$, e.g., as in~\cite{Boyd_2004}. 
Time-varying  connections such as randomized 
interconnections also can lead to a faster response,  e.g.,~\cite{Carli_08}. Moreover,  connectivity enhancements have been proposed for  jointly-connected networks
~\cite{Zongli_Shize_19}. 
Nevertheless,  
when such time-variations in the graph structure or selection of the graph Laplacian $K$ are not feasible (e.g., when a given structure has to be used), 
the range of acceptable update gain $\gamma$, e.g., due to input bounds, can limit the response speed.
Finally, although speeding up the response leads to smaller loss of cohesion, the response-time delays are still present  in the faster response. 
Cohesion, normalized by the settling time, does not necessarily improve. 
The lack of cohesion, even with faster response and potential limits due to actuator constraints,  motivates 
the current effort to improve  cohesion without the typical emphasis on increasing the response speed.

Ideally, for cohesion, all agents should be directly connected to the source. 
Then, every agent has immediate 
access to the desired response, i.e.,  $z_s$. 
However, this requires  broadcasting the source information across spatially distant neighbors that might not be feasible 
in large networks. Moreover, such broadcasting might not be preferred in the presence of adversaries since they could then infer the intent of the network. 
The cohesion problem addressed here is to achieve a uniform response across the network (to changes in the source), and 
each agent uses information from its neighbors without requiring additional  knowledge  about the overall network connections. 

Previous work has shown that cohesion can be improved by using  derivative information 
from the neighbors of each agent, see~\cite{Ren_07}. In such a setting, 
the control input for an agent $i$, contains derivative information from its neighbors $N_i$, which in turn 
depend on derivative information from other neighbors $k \notin N_i$.  Therefore, each agent $i$ cannot independently compute its update $\dot{z}_i$ by only knowing 
information about its immediate neighbors $N_i$. 
It is shown in this article that the proposed, delayed self reinforcement (DSR), effectively 
approximates the derivative information  from the neighbors needed for response cohesion. Such use of delayed information has been used  in artificial neural networks for improving gradient-based learning  algorithms~\cite{Rumelhart_86,QIAN1999145}. 
Similar use of delayed information can improve network response speeds  under update-bandwidth limitations for 
discrete-time multi-agent systems, e.g., 
as shown in ~\cite{Devasia_2018_JDSMC}. The novelty in the current work is the use of 
DSR to  improve cohesion using  already-available information from a given network, without requiring network-connectivity modifications  
or increases  in the response speed.

The main contribution of this work is the development of  stability conditions for the proposed delayed self-reinforcement (DSR) approach. 
The delay in the implementation turns  the dynamics of the networked system into a delay-differential-equation (DDE).
Note that DDEs have been well studied in the past, e.g., ~\cite{Miranker_62,Bellman_Cooke_63}, and numerical methods are available using the 
Lambert W function~\cite{Corless1996}  to evaluate the stability of DDEs, e.g. see~\cite{Asl_Ulsoy_2003,Yi_Ulsoy_2010}. 
For example, derivative control, used to improve robustness of single-input-single-output  systems, can be implemented using delay-based approximation as in~\cite{Ulsoy_2015}, and 
stability can be inferred using the Lambert W function. 
Approaches have also been developed to find the range of time-delays under which  stability is maintained for a DDE, e.g.,~\cite{Chen_IEEE_95,Nejat_02,Nejat_05,Qiao_Siplahi_16}.
Nevertheless,  it is challenging to develop general stability conditions for DDEs. 
For special cases, 
e.g., when the matrices involved in the DDEs are symmetric (which corresponds to the underlying 
graph associated with the Laplacian $K$ being undirected in the current application)  
stability conditions can be developed, e.g., as in~\cite{Brayton_67}. 
The current paper develops  generalized stability conditions for the DDE associated with the DSR approach 
for, both, directed and undirected graphs. 
These stability conditions are developed by exploiting the  graph structure of the network and the results 
depend  on the eigenvalues of the associated graph Laplacian $K$. 
In this sense it extends Brayton's 
stability results for DDEs ~\cite{Brayton_67} to  the more general case  with non-symmetric matrices, which  can be applied to directed graphs. 
Moreover,  the article shows that  when the eigenvalues of the Laplacian $K$ are real, e.g., for undirected graphs or  directed but topologically ordered sub-graphs (defined later 
in this article), (i)~ the stability conditions 
developed in this article for the  network with DSR reduce to the results for 
scalar DDEs from~\cite{Hayes_50}, and 
(ii)~the  proposed DSR approach is stable independent of the delay. 
Lastly, the proposed DSR approach is applicable to cases when the agent dynamics 
is heterogenous and higher order, but with the same relative degree.

\vspace{-0.1in}
\section{Cohesive-response problem}
\vspace{-0.1in}
The network dynamics is defined using a graph representation in this section. Then, the cohesion in the response dynamics is quantified and the problem of 
improving the cohesion is posed.

\vspace{-0.1in}\subsection{Graph-based response dynamics}
  \vspace*{-0.1in}\noindent
Let the connectivity of the agents be represented by 
a directed graph (digraph) ${\mathcal{G}} = \left({\mathcal{V}}, {\mathcal{E}}\right)$, e.g., 
as defined in~\cite{Olfati_Murray_07}, with agents represented by nodes $ {\mathcal{V}}= \left\{ 1, 2, \hdots, {n\!+\!1} \right\}$, $n>1$ 
and edges $ {\mathcal{E}}   \subseteq {\mathcal{V}} \times {\mathcal{V}} $, where the 
neighbors of the agent $i$ are represented by the set  $N_i = \{ k \in {\mathcal{V}} , k \ne i:  (k,i) \in {\mathcal{E}} \}$. 
Node $s$, which is assumed, without loss of 
generality,  to be 
the last node, 
represents the   desired response,  $z_s$.
The terms
 $l_{ik}$ of the $(n+1)\times(n+1)$ Laplacian $L$  of the graph ${\mathcal{G}}$ are real and given by 
\begin{align}
\label{eq_laplacian_defn}
l_{ik} & =   \left\{ 
\begin{array}{ll}
-w_{ik}, 	& {\mbox{if}} ~ k \in N_i \\
\sum_{m=1}^{n+1} w_{im}, & {\mbox{if}} ~ k = i,  \\
0 & {\mbox{otherwise, }}
\end{array}  
\right.
\end{align}
where the weights $w_{ik}$ are positive if $k \in N_i$ and   zero otherwise. 
The  dynamics  for the non-source agents $Z$ (with each agent state-component given by $z_i$), 
represented by the graph ${\mathcal{G}}\!\setminus\!s$, can be written in matrix form as 
\begin{align}
\frac{d{Z}}{dt}  (t)  = \dot{Z} (t)  & =  U    = -K Z (t) +B  z_s (t), 
\label{system_non_source}
\end{align}
similar to Eq.~\eqref{intro_sys_eq}, 
where $U$ is the input to the agents. 
The $n \times n$ matrix $K$ (the pinned Laplacian) is obtained by removing the row and column associated with the source node $n+1$ through 
the following partitioning of the graph Laplacian $L$, i.e., 
\begin{align}
\label{eq_K_eigenvector}
L & = 
\left[
\begin{array}{c|c}
K  &  -B  \\
 \hline
  \star_{1 \times n} & \star_{1 \times 1} 
\end{array}
\right] 
\end{align}
with  $B$  an  $n \times 1$ input matrix, 
$B  = [ w_{1,s}, w_{2,s}, \hdots, w_{n,s}]^T~= [ B_{1}, B_{2}, \hdots, B_{n}]^T $.

\subsection{Graph properties}
\vspace*{-0.2in}\noindent
Some standard graph properties (needed later), resulting from the following assumption,  are described below.
\begin{assumption}[Connected to source node]
\label{assumption_digraph_properties}
The digraph ${\mathcal{G}}$ is assumed  to have a directed path from the source node $s$ to any  node $ i \in {\mathcal{V}} \setminus \!s$. 
\end{assumption}

From Assumption~\ref{assumption_digraph_properties} and the  Matrix-Tree Theorem in~\cite{tuttle_graph}  the pinned Laplacian $K$ of the graph without the source node $s$ is invertible, i.e., 
$\det{(K)} \ne 0$. 
The eigenvalues $\left\{ \lambda_{K,i} \right\}_{i=1}^{n}$ of pinned Laplacian $K$  have strictly-positive, real parts, i.e., 
\begin{align}
\label{eq_real_part_K}
{{\mathcal{R}}e} \left({\lambda_{K,i} }\right) & > 0, 
\end{align}
and therefore 
the negative of the pinned Laplacian (i.e., $-K$) is Hurwitz, with eigenvalues on the open left half of the complex plane. 
This follows from the Gershgorin theorem since all the eigenvalues of  the pinned Laplacian $K$ must lie in one of circles centered at $l_{ii} >0$ with radius $ l_{ii}-w_{is}  \in [0, l_{ii}] $  from definition of $l_{ii}$ in Eq.~\eqref{eq_laplacian_defn} and  $w_{is}\ge 0$.  
Given the invertibility of pinned Laplacian $K$, the   
eigenvalues of the pinned Laplacian $K$ cannot be at the origin, 
and therefore the eigenvalues must have strictly positive real parts (from the Gershgorin theorem condition of being inside the circles which are on the right hand side of the complex place except for the origin). 

The product of the inverse of the pinned Laplacian $K$ with $B$ leads to a  $n \times 1$ vector of  ones, 
i.e., 
$K ^{-1} B  = {\textbf{1}}_n$, 
which follows from the partitioning in Eq.~\eqref{eq_K_eigenvector}, and invertibility of $K$ since 
the $(n+1) \times 1$ vector of ones  ${\textbf{1}}_{n+1}= [1, \hdots, 1]^T$  is a right eigenvector of the Laplacian $L$ with eigenvalue $0$,  i.e., 
$
L{\textbf{1}}_{n+1} 
= 0  {\textbf{1}}_{n+1}, 
$
resulting in 
\begin{align}
\label{eq_KinvtimesB_2}
K {\textbf{1}}_n  & =  B.
\end{align}

\vspace{-0.1in}
\subsection{Quantifying cohesion}
\vspace*{-0.1in}\noindent
Lack of cohesion is quantified in terms of the 
deviations $\Delta$  in the responses between agents for a  step change in the  source $z_s$ 
from $z_s(0)=0 $ at time $t=0$ to $z_s(t) = z_d \ne 0$ for time $t> 0$. 
The  response of the non-source agents, i.e., solution to Eq.~\eqref{system_non_source}, 
can be written as  
\begin{align}
Z(t) &  = e^{-K t} Z(0) +(-K)^{-1} \left[  e^{-K t} - I  \right] B z_d, 
\label{Eq_controlled_gen_soln}
\end{align}
which simplifies to 
\begin{align}
Z(t) &  = (-K)^{-1} \left[  e^{-K t} - I  \right] B z_d
\label{Eq_consensus_soln}
\end{align}
if the initial state  $Z(0)$ is at consensus, i.e., $Z(0)=0$.
Note that the exponent $e^{-K t} \rightarrow 0 $ as time increases since the negative of the pinned Laplacian $-K $ is Hurwitz. 
Therefore,    from Eq.~\eqref{eq_KinvtimesB_2}, the response  $Z(t)$,  of the non-source agents, exponentially reaches 
the desired value $z_d$ as time $t$ increases, i.e, 
\begin{align}
Z (t)  &    \rightarrow z_d {\textbf{1}}_n  ~~{\mbox{as}}~~  t \rightarrow \infty.
\label{system_non_source_stability}
\end{align}
\noindent
The lack of cohesion can be quantified in terms of the 
deviations $\Delta$  in the response 
as 
\begin{align}
\Delta &  = \frac{1}{z_d } \int_0^{T_s}  \left| Z(t) - \overline{z}(t)  {\textbf{1}}_n \right|_1 dt , 
\label{Eq_hl_cohesion}
\end{align}
where 
$T_s$ is the settling time, i.e., the time by which all agent responses $Z$ reach and stay within $2\%$ of the final value $z_d$,  
$ \overline{z}$ is the average value of the state $Z$, over all individual agent state-components $z_i$, i.e., 
\begin{align}
 \overline{z}(t)  & = \frac{1}{n} \sum_{i=1}^{n} z_i(t), 
\label{Eq_hl_cohesion_2}
\end{align}
and 
 $|  \cdot |_1  $ is the standard vector 1-norm, 
$| \hat{Z} |_1  = \sum_{i=1}^{n} | \hat{z}_i|$ for any vector $\hat{Z}$. 
A normalized measure ${\Delta}^*$  that removes the effect of the response speed is obtained by 
dividing the expression in Eq.~\eqref{Eq_hl_cohesion} with the settling time $T_s$ as 
\begin{align}
{\Delta}^* &  =   \frac{\Delta}{T_s} .
\label{Eq_norm_hl_cohesion}
\end{align}
Note that the system's transient response is more cohesive  if the normalized deviation ${\Delta}^* $ is small.

\vspace{-0.1in}\subsection{Problem: reduce normalized deviation ${\Delta}^*$}
\vspace*{-0.1in}\noindent
The research problem  is to improve cohesion (i.e.,  to reduce the normalized deviation ${\Delta}^*$)  
without changing the network connectivity or access to the information source, i.e., without changing the network graph  ${\mathcal{G}} $.

\section{Proposed approach}

\vspace{-0.1in}\subsection{Ideal cohesive  dynamics}
\vspace*{-0.1in}\noindent
 If each non-source agent can have instantaneous access to the source $z_s$ 
 \begin{align}
 \dot{z}_i (t)  &  = -  \alpha {z}_i(t)  + \alpha z_s(t) , 
\label{system_limit_single}
\end{align}
then with the same initial condition, the response of all the agents would be cohesive. 
In particular,  the entire system will respond to a step input, 
$z_s(0)=0 $ at time $t=0$ to $z_s(t) = z_d \ne 0$ for time $t> 0$ with a zero initial state $z_i(0)=0$,  
in a cohesive  manner.  Moreover, each agent state $z_i$ will have the same 
settling time  (to reach and stay within 2\% of the final value) 
\begin{align}
T_s  =  \frac{4}{   \alpha  } , 
\label{Settling_time}
\end{align}
provided, for stability,  
$ \alpha > 0$. 
Note that the parameter $\alpha$ can be used to adjust the 
overall speed of the response of each agent. 
In a vector form, this ideal cohesive  dynamics can be written as 
\begin{align}
 \dot{Z} (t)  &  = -  \alpha Z (t)  +  \alpha {\textbf{1}}_n  z_s(t) .
\label{system_limit}
\end{align}
\noindent
Multiplying both sides of Eq.~\eqref{system_limit} by $\beta K$  (where $\beta > 0$) and using Eq.~\eqref{eq_KinvtimesB_2} to replace $K {\textbf{1}}_n$ results in 
\begin{equation}
\begin{aligned}
 \beta K \dot{Z} (t)  &  = -   \alpha  \beta K  Z (t)  +  \alpha \beta K {\textbf{1}}_n  z_s(t)  
                         ~~ =  -    \alpha  \beta K Z (t)  + \alpha \beta    B  z_s (t), 
\end{aligned}
\label{system_imit_mod1}
\end{equation}
and therefore, by adding $\dot{Z} $ on both sides,  the ideal cohesive  dynamics can be rewritten  as 
\begin{align}
\dot{Z} (t)  &  = -\alpha \beta K Z (t)  + \alpha\beta    B  z_s (t)  + \left[ I - \beta K \right] \dot{Z} (t ) .
\label{system_R}
\end{align}

\vspace{0.05in}
\begin{rem}[Network connectivity and improved cohesion]
\label{Rem_improved_hl_cohesion}
The ideal cohesive  dynamics  in 
Eq.~\eqref{system_R} is found by exploiting the network connectivity (i.e., invertibility of the pinned Laplacian, and convergence to consensus), which enables the replacement of 
$K {\textbf{1}}_n $ by $B$ in Eq.~\eqref{system_imit_mod1}. As a result, 
all agents have the same time-trajectory solution as in Eq.~\eqref{system_limit} 
(provided the system is initially synchronized)
even with rapid changes in the source $z_s$. 
Therefore, the normalized deviation 
in \eqref{Eq_norm_hl_cohesion} 
is zero with the ideally cohesive  dynamics in 
Eq.~\eqref{system_R}, i.e., 
$$ {\Delta}^* =0.   $$
In this sense, the DSR approach (presented below) accounts for network issues such as potential redundancy 
of information obtained from neighbors. 
 \hfill \qed
\end{rem}

The measure of cohesion ${\Delta}^*$ does not capture the convergence rate to the 
final value, which depends on (and can be adjusted by the selection of) the parameter $\alpha$ 
in Eq.~\eqref{system_limit}.

\vspace{-0.1in}\subsection{Delay-based derivative}
The idealized control law in Eq.~\eqref{system_R}  is 
implemented in the following with  delayed self reinforcement (DSR). 
The input for an agent $i$, from the right hand side of the expression in Eq.~\eqref{system_R}, contains derivative information from its neighbors $N_i$, which in turn 
depends on other agents $k$ that might not be a neighbor of agent $i$, i.e., $k \notin N_i$.  Therefore, it is challenging to compute, 
and  implement the  ideal cohesive  dynamics in Eq.~\eqref{system_R}. A delay-based implementation 
of the derivative is discussed below.

Consider the approximate implementation of the derivative $ \dot{Z} (t )$  
in right-hand-side of Eq.~\eqref{system_R} by using a delay, $\tau >  0,$  
as
\begin{align} 
\dot{Z} (t)  & \approx \frac{ Z(t)  - Z(t-\tau)}{\tau}  .
\label{system_DR_delayed_derivative}
\end{align}

\vspace{-0.2in}
\begin{rem}[Delayed derivative acts as a filter]
\label{rem_Delayed_derivative_acts_filter}
The  gain associated with the 
standard time derivative $\frac{d}{dt}$, with Laplace transform $s$, grows linearly with frequency. In contrast, the 
 gain  associated with the delay-based approximate derivative   $\frac{1 }{\tau} K [ Z(t) -  Z(t -\tau)] $ in  Eq.~\eqref{system_DR_delayed_derivative} , with  Laplace transform
 $ \frac{ 1  - e^{-\tau s}}{\tau} $  
is  bounded by $  \frac{2}{\tau}$ over all frequency $s = j\omega.$  Thus, the approximated 
derivative acts as a filtered derivative at higher frequencies, especially when 
the delay is chosen based on the overall settling time in Eq.~\eqref{Settling_time}, say 
\begin{align}
\tau  & = \frac{T_s}{100}  =  \frac{4}{100 \alpha} .
\label{eq_tau_Ts_link}
  \hfill \qed
 \end{align}
\end{rem}

\vspace{-0.1in}
Substituting for the derivative $ \dot{Z} (s)$  in right-hand-side of Eq.~\eqref{system_R}
with the expression in Eq.~\eqref{system_DR_delayed_derivative} 
results in a modified system described by the following delay-differential-equation (DDE)
\begin{align}
\dot{Z} (t)  &  
= U   = A Z(t) + A_d Z(t -\tau)   + B_d z_s (t) 
\label{system_DR}
\end{align}
where $U$ is the input  
and 
\begin{equation}
\begin{aligned}
A & =  -\alpha \beta  K   +\frac{1}{\tau}\left[ I -  \beta K    \right] \\
A_d & = - \frac{1}{\tau}\left[  I -   \beta K \right]  \\
B_d  & = \alpha \beta  B .
\label{system_DR_matrices}
\end{aligned}
\end{equation}

\vspace{0.05in}
\vspace{-0.1in}\subsection{Cohesive tracking}
\label{subsectio_tracking}

\vspace*{-0.2in}\noindent

The same delay-based implementation could be used to achieve cohesive  tracking (where all agents have the same response if the initial conditions 
for time $t \in [-\tau, 0]$ are the same for all agents) 
when the desired response $z_s$ is differentiable and the 
system dynamics in Eq.~\eqref{system_limit} is modified to 
\begin{align} 
 \dot{Z} (t)  &  = -  \alpha Z (t)  +  {\textbf{1}}_n \left[  \alpha z_s(t)   + \dot{z}_s \right] 
\label{system_sync_tracking}
\end{align}
since the tracking error $e_i = z_i - z_s$ of each non-source agent $i$, due to initial condition errors,  converges to zero because 
\begin{align} 
 \dot{e}_i (t)  &  = -  \alpha e_i (t) . 
\label{system_sync_tracking_error}
\end{align}
As in Eq.~\eqref{system_R}, the tracking dynamics in Eq.~\eqref{system_sync_tracking} can be rewritten as 
\begin{align}
\dot{Z} (t)  &  = -\alpha \beta K Z (t)   + \left[ I - \beta K \right] \dot{Z} (t )
+ \beta    B \left[  \alpha z_s(t)   + \dot{z}_s \right] .
\label{system_R_tracking}
\end{align}

\begin{rem}[Prior use of derivative information]
The use of derivative information for trajectory tracking in Eq.~\eqref{system_R_tracking} is similar to previous work that uses such derivative information, e.g.,~\cite{Ren_07}. 
In particular, the above tracking dynamics in Eq.~\eqref{system_R_tracking} can be rewritten 
for an individual agent $i$   as  
\begin{align}
  \dot{z}_i  & =  \frac{1}{\eta}  \sum_{k=1}^n w_{i,k}  \left( \dot{z}_k (t)    -\alpha \left[{z}_i(t) -{z}_k (t)    \right]  \right) 
~ +   \frac{1}{\eta}  w_{i,s} \left(  {  \dot{z}_s  -\alpha \left[{z}_i (t) -{z}_s (t) \right] } \right) , 
\label{system_imit_mod_tracking_3}
\end{align}
where 
$ \eta   =   w_{i,s} +   \sum_{k=1}^n w_{i,k}  $
that is similar to the derivative-based control law in Eq.~(7) of~\cite{Ren_07}.  
However, such a derivative-based approach is difficult to implement since  derivatives appears on both sides of the equation. This implies that  neighbors need to know, simultaneously, each others time derivatives $  \dot{z}$ to compute 
their own time derivatives. 
 \hfill \qed
\end{rem}

To avoid the need to know the time derivatives $\dot{z}_k$ to compute the time derivative $  \dot{z}_i $ in Eq.~\eqref{system_imit_mod_tracking_3}, a delay-based implementation, as in Eq.~\eqref{system_DR}, is given  by the following DDE 
\begin{align}
\dot{Z} (t)  &  
= A Z(t) + A_d Z(t -\tau)   + \left[  \alpha \beta B  z_s (t)  + \beta B \dot{z}_s \right] \\[0.3em]
& = -\alpha \beta  K  Z(t)   +\frac{1}{\tau}[ Z(t) -  Z(t -\tau)]       
  - \frac{\beta }{\tau} K [ Z(t) -  Z(t -\tau)]   + \left[  \alpha \beta B  z_s (t)  + \beta B \dot{z}_s \right] .
\label{system_DR_tracking}
\end{align}

\begin{rem}[Connection to  optimization algorithms]
\label{rem_connection_to_learning_algorithms}
The proposed DSR  has a similar form 
as reinforcement terms used in 
gradient-based,  optimization algorithms, e.g.,  $[ Z(t) -  Z(t -\tau)]  $ in the second term  of the control law in   Eq.~\eqref{system_DR_tracking} 
is referred to as the  momentum term~\cite{Rumelhart_86} and  
 $K [ Z(t) -  Z(t -\tau)] $ in the third term is referred to as the Nesterov 
term or the acceleration term~\cite{Nesterov_83}.  Recently, for discrete-time systems, 
the use of the momentum term alone (without the Nesterov term) to improve  
the response speed  of swarms and networks under update-bandwidth limits 
has been shown in~\cite{Devasia_2018_JDSMC,Devasia_2019_IJC}, and the use of the Nesterov term alone (without 
the momentum term) has been shown to have a faster rate of convergence to consensus in~\cite{Cao_ren_2010,Moradian_19}, as well as a linear rate of convergence in~\cite{Bu_2018}. 
More recently, both the momentum and Nesterov 
terms, in the same ratio, has been shown to improve convergence rate~\cite{Devasia_ICPS_2019}. 
Thus, the improved-cohesion argument in the current work provides a rationale for prior Nesterov-type accelerated optimization methods~\cite{Rumelhart_86,Nesterov_83}, and generalizes such accelerated methods (currently available only for agents with first-order dynamics) to agents with higher-order dynamics. 
\hfill \qed
\end{rem}

\subsection{Network information needed for DSR}
\begin{figure}[!ht]
\begin{center}
\includegraphics[width=.5\columnwidth]{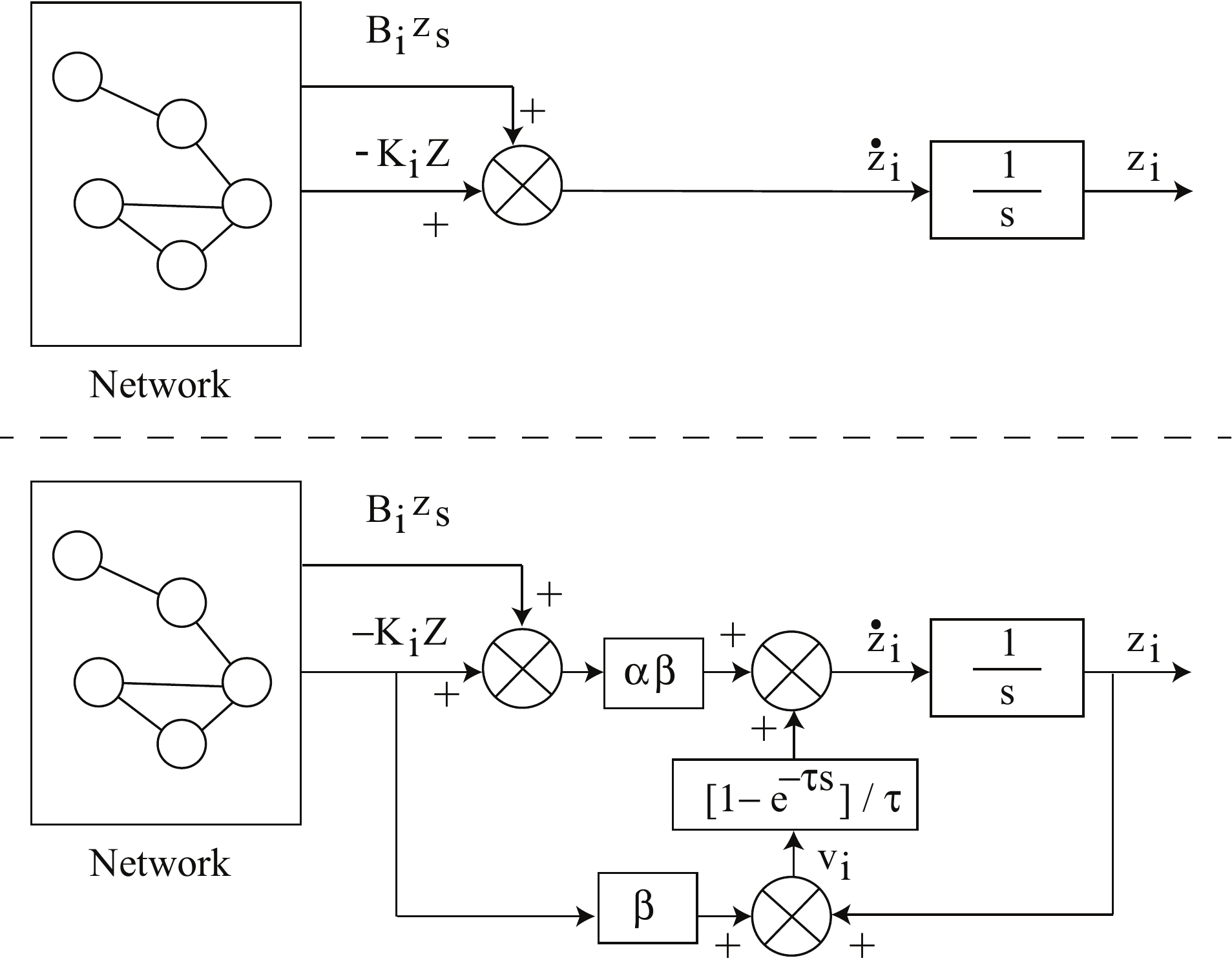}
\vspace{0.01in}
\caption{(Top) Without DSR: dynamics of agent $i$ for original  networked system without delayed reinforcement. 
(Bottom) With DSR: modified dynamics of agent $i$  with delayed reinforcement using the same network information  ($K_i Z $ and $B_i z_s $, where
$B_i$ and $K_i$ are the $i^{th}$ rows of matrices $B$ and $K$). 
}
\label{fig_1_control_implementation}
\end{center}
\end{figure}

   \vspace*{-0.1in}\noindent
The computation of the input $U$ to the individual agents, on the right-hand-side of Eq.~\eqref{system_DR}, does not require additional information from the 
network.  The DSR input $U$ is reinforced with a 
delayed-version of already-available information.  
For example, the $i^{th}$ agent dynamics in Eq.~\eqref{system_limit_single} is modified, 
according to Eq.~\eqref{system_R}, as 
\begin{align}
\dot{z}_i (s)  &  = -   \alpha\beta   K_i Z (s)     +\alpha\beta  B_i  z_s (s) +  
\left( \frac{1 - e^{-\tau s}  }{\tau}\right) v_i(s)  , 
\label{system_non_source_i_mod}
\end{align}
where
$B_i$ and $K_i$ are the $i^{th}$ rows of matrices $B$ and $K$, and  the additional input term $v_i$ is  computed 
without modifying the network structure $K$, 
\begin{align}
v_i( \cdot ) =    z_i ( \cdot)   -   \beta K_i  Z  ( \cdot )   , 
\label{system_non_source_i_2_mod}
\end{align}
as illustrated in Fig.~\ref{fig_1_control_implementation}.

\suppressfloats

 \vspace{-0.1in}\subsection{Cohesiveness with DSR}
   \vspace*{-0.1in}\noindent
The following lemma shows that 
the  DSR approach leads to solutions $Z$ of Eq.~\eqref{system_DR_tracking} that are 
close to the ideal cohesive dynamics in Eq.~\eqref{system_R_tracking}, if either the dominant dynamics  is sufficiently slow (magnitude of  $\ddot{Z}$ is small) 
or if the delay $\tau$ is sufficiently small.

\begin{Lemma}[DSR and ideal cohesive dynamics]
\label{Lemma_DSR_cohesiveness}  
~~~ Let the  source $z_s$ be sufficiently smooth in a finite time interval ${\mathcal{I}_t} = [t_1, t_2]$.  
Then, 
solutions $Z$ of the DSR Eq.~\eqref{system_DR_tracking} are close to the ideal 
cohesive solution $Z^*$ of Eq.~\eqref{system_R_tracking}, with the same synchronized initial 
conditions, provided 
the product of the time delay $\tau$ and the  maximum acceleration $\ddot{Z}$ 
over the interval ${\mathcal{I}_t}$ is small. 
Formally,  the deviation $E_Z  = Z - Z^*$ satisfies 
\begin{align}
\max_{t \in {\mathcal{I}_t}} \| E_Z (t) \|_\infty   \rightarrow 0,   \quad {\mbox{as}}  
\quad   \max_{t \in {\mathcal{I}_t}}  \|\tau \ddot{Z}( t) \|_{\infty} \rightarrow 0 .
\end{align}

\end{Lemma}

\noindent
{\textbf{Proof}}~
Solutions $Z^*$ and $Z$ to Eqs.~\eqref{system_R_tracking} and  \eqref{system_DR_tracking}, respectively, are twice differentiable in the time interval $(t_1, t_2)$ if the source $z_s$ is sufficiently smooth. 
From Taylor's theorem, given the differentiability of  $Z$,  
\begin{align}
Z (t-\tau) = Z(t) + [-\tau] \dot{Z}(t)  + \tau^2 H(t),  
\end{align}
where $H$ is bound by the maximum agent acceleration, i.e.,  
\begin{align}
\| H(t) \|_\infty  ~ \le  ~ \max_{1\le i \le n, t \in (t_1, t_2)}  \frac{1}{2}  | \ddot{z}_i (t) |  ~= \overline{H} .
\end{align}
Substituting for the approximate derivative  
$ \frac{Z(t)  - Z (t-\tau) }{\tau} = \dot{Z}(t)  -   \tau H(t)  $
in Eq.~\eqref{system_DR_tracking}, and reversing the arguments   
 from Eqs.~\eqref{system_limit} to \eqref{system_R}, results in 
\begin{align}
 \dot{Z} (t)  &  = -  \alpha Z (t)  +  {\textbf{1}}_n \left[  \alpha z_s(t)   + \dot{z}_s \right] 
 +  \frac{\tau}{\beta} K^{-1}  \left[ I - \beta K \right]  H(t) .
\label{system_sync_tracking_star}
\end{align}
Let $Z^*$ be a solution to ideal cohesive dynamics in Eq.~\eqref{system_sync_tracking},
\begin{align} 
 \dot{Z}^* (t)  &  = -  \alpha Z^* (t)  +  {\textbf{1}}_n \left[  \alpha z_s(t)   + \dot{z}_s \right] .
\label{system_sync_tracking_star_2}
\end{align}
Then,  the dynamics of the deviation $E_z  = Z - Z^*$ between the two solutions $Z$ and $Z^*$ can be found 
subtracting Eq.~\eqref{system_sync_tracking_star_2} from Eq.~\eqref{system_sync_tracking_star} to obtain 
\begin{align}
 \dot{E}_Z (t)  &  = -  \alpha E_Z (t)   +  \frac{\tau}{\beta} K^{-1}  \left[ I - \beta K \right]  H(t), 
\label{system_sync_tracking_error}
\end{align}
which is bounded-input bounded-output stable. 
If the initial conditions  at time $t_1$ are the same, i.e., $E_Z(t_1)=0$  then the result follows since the 
deviation $E_Z$ tends to zero as 
the maximum magnitude of $ \tau H(t)$ (which is not bigger  than $ \tau \overline{H}$)  tends to zero. 
\hfill \qed

\begin{rem}[DSR cohesiveness]
\label{rem_DSR_cohesiveness}
From the above lemma, the use of DSR can lead to solutions close to the ideal  cohesive dynamics resulting in a smaller cohesion error   $\Delta^*$, provided 
the  product of the maximum magnitude $\ddot{Z}$ and time delay $\tau$ is sufficiently small. 
However, without DSR, such a reduction is not possible for a general network. Moreover,  
the DSR approach (with $\beta=1$ and the derivative term $\dot{Z}$ set to zero 
on the right hand side of Eq.~\eqref{system_R}) can 
perform as well as the case without the DSR. Therefore, in general,  the use of 
DSR can improve cohesiveness when compared to the case without DSR. 
\end{rem}

\section{Stability analysis}
\vspace*{-0.1in}\noindent
This section begins with a numerical check for stability, followed by 
conditions on the parameters ($\alpha, \beta, \tau$) for stability of the 
DSR approach.

 \vspace{-0.1in}\subsection{Eigenvalues of pinned Laplacian $K$ and stability}
 \vspace*{-0.1in}\noindent
With general matrices $A$, $A_d$ in DDE Eq.~\eqref{system_DR} it is difficult to relate the stability of the 
DDE to the eigenvalues of the two matrices. However, given the special structure of $A$, $A_d$ in the current DDE, the 
stability of the DDE can be related to the eigenvalues 
$\left\{ \lambda_{K,i} \right\}_{i=1}^{n}$ of the pinned Laplacian $K$, as shown below.

\begin{Lemma}[Stability of DDE]
\label{Lemma_DDE_eigenvalues}
The DDE system in Eq.~\eqref{system_DR}   is   exponentially stable 
if and only if the roots 
$s_i$ of 
\begin{align}
s  -  \lambda_{i} -   \lambda_{d,i} e^{-s \tau}  & = 0  
\label{stability_characteristic_eq_new_0}
\end{align}
have negative real part, i.e., 
\begin{align}
{{\mathcal{R}}e\{s_i\}}  <   0   
\label{eq_cond_negative_real_part}
\end{align} 
for all integers $1\le i \le n$ where 
\begin{equation}
\begin{aligned}
\lambda_{i} & =  - \alpha\beta \lambda_{K,i}  +\frac{1}{\tau}\left[ 1  - \beta \lambda_{K,i} \right]  \\
\lambda_{d,i} & = - \frac{1}{\tau}\left[1  - \beta \lambda_{K,i}\right], 
\label{system_eigenvalues_DR_matrices}
\end{aligned} 
\end{equation}
and
$\left\{ \lambda_{K,i} \right\}_{i=1}^{n}$  are eigenvalues of the pinned Laplacian $K$. 

\end{Lemma}

\noindent
{\textbf{Proof}}~
To begin, the DDE~\eqref{system_DR}  is converted into a Jordan form. 
Let the pinned Laplacian $K$ be similar to the diagonal
matrix $K_J$ in the Jordan form, where  
\begin{equation}
K_J = P_K^{-1} K P_K
\label{Jordan_K_eq}
\end{equation}
and
the diagonal  terms of matrix $K_J$ 
are the (potentially complex-valued)  eigenvalues $\left\{ \lambda_{K,i} \right\}_{i=1}^{n}$   of matrix $K$~\cite{Ortega}. 
Note that the  multiplicity of each eigenvalue $ \lambda_{K,i} $ can be more than one. 
The invertible (potentially complex-valued) matrix $P_K$ also transforms $A , A_d $ in the DDE Eq.~\eqref{system_DR} into Jordan-like forms $A_{J}, A_{d,J}$  with  
(potentially complex-valued)  diagonal terms $ \lambda_{i} ,\lambda_{d,i} $ described in Eq.~\eqref{system_eigenvalues_DR_matrices} since, from Eqs.~\eqref{system_DR_matrices} and \eqref{Jordan_K_eq}, 
\begin{equation}
\begin{aligned}
A_{J} ~~= P_K^{-1}A P_K & =  -\alpha\beta   K_J    +\frac{1}{\tau}\left[ I - \beta K_J   \right] \\
A_{d,J}~~=  P_K^{-1}A_d P_K & = -\frac{1}{\tau}\left[   I - \beta K_J  \right]  .
\label{system_DR_matrices_temp}
\end{aligned}
\end{equation}
Then,  
setting the input $z_s$ to zero 
and 
changing the coordinates in  the DDE Eq.~\eqref{system_DR}  to 
$Z(t) ~ = P_K Z_J(t)$, 
and pre-multiplying  by $P_K^{-1}$  on both sides results in 
\begin{align}
\dot{Z}_J (s)  &  
= A_J Z_J(s) +e^{-s\tau} A_{d,J} Z_J(s) . 
\label{system_DR_new_coordinates}
\end{align}
The stability of the DDE~\eqref{system_DR},  is equivalent to the stability of the DDE~\eqref{system_DR_new_coordinates} in the new coordinates $Z_J$. 
In particular, system is exponentially stable if the roots $s = \lambda_{DDE}$ of the 
characteristic equation, 
\begin{equation}
\det \left|  s I - A_J -A_{d,J} e^{-s\tau}       \right|  = 0, 
\label{characteristic_eq_Miranker}
\end{equation}
satisfy 
~\cite{Miranker_62,Yamamoto_91}
\begin{align}
\sup  {{\mathcal{R}}e}{(\lambda_{DDE})}  <  -\gamma_\lambda  < 0   ~ .
\label{stability_characteristic_eq_max_real}
\end{align} 
Since $ A_{J}$ and $ A_{d,J}$ have 
triangular (Jordan) forms, the characteristic Eq.~\eqref{characteristic_eq_Miranker} can be rewritten,  by considering the diagonal terms), as 
\begin{align}
\prod_{i=1}^{n}  \left[  s -  \lambda_{i}    - \lambda_{d,i} e^{-s \tau}  \right]  = 0, 
\label{stability_characteristic_eq}
\end{align} 
whose roots $s = \lambda_{DDE}$  are the same as roots $s_i$ of Eq.~\eqref{stability_characteristic_eq_new_0} of the lemma.

Finally, properties of analytic functions, 
can be used to show that roots $s_i$  satisfying the negative real part condition  in Eq.~\eqref{eq_cond_negative_real_part} also satisfy  the more stricter stability condition in Eq.~\eqref{stability_characteristic_eq_max_real} with  $\lambda_{DDE} = s_i$.  
If the roots  $\lambda_{DDE}$ have negative real parts but there is no $\gamma_\lambda> 0$ satisfying Eq.~\eqref{stability_characteristic_eq_max_real}, then there is an infinite number of roots $\lambda_{DDE}$ arbitrarily  close to the imaginary axis ${\mathcal{R}}e\{s\} = 0$. This follows by considering roots to the right of the sequence of lines ${\mathcal{R}}e\{s\} = -1/N$ that are getting closer to the imaginary axis as $N$ increases.  One can find a subsequence of these lines such that there is a sequence of distinct roots to the right of each line. Note that the roots close to the imaginary axis have finite magnitude.  There exists  constants $\overline{s}, \bar{N} $ such that 
there are no roots   close to the imaginary axis (to the right of the line 
${\mathcal{R}}e\{s\} = -1/\bar{N}$) with $|s| > \overline{s} $
because the portion of the characteristic Eq~\eqref{stability_characteristic_eq_new_0} is dominated by the first term $s$ and both the other terms 
$ \lambda_{i} $ and $ \lambda_{d,i}e^{-s \tau} $ are bounded on (and close to) the imaginary axis.
So there are infinite roots residing in each bounded region satisfying $ |s|  \le \overline{s} $ and $ {\mathcal{R}}e\{s\} > -1/N$, which is not possible since Eq.~\eqref{stability_characteristic_eq_new_0} is analytic 
and can only have a finite number of zeros in any bounded region. 

The necessity of the  negative-real-part condition in  Eq.~\eqref{eq_cond_negative_real_part} 
follows since the DDE has solutions with terms of the form $e^{s_i t} Z_{s,i}$, where $Z_{s,i}$ 
is the eigenvector associated with 
eigenvalue $s_i$.
 \hfill \qed

 \vspace{0.1in}
  \subsection{Numerical check for stability}
  \vspace*{-0.1in}\noindent
Stability of the DSR could be checked numerically. 
For example, 
the roots of the characteristic Eq.~\eqref{stability_characteristic_eq} are composed of the roots $s_i$ in Eq.~\eqref{stability_characteristic_eq_new_0} of the individual terms forming the 
product in the characteristic  Eq.~\eqref{stability_characteristic_eq}, i.e., solutions to $s_i   -  \lambda_{i}    -  \lambda_{d,i} e^{-s_i \tau}   = 0  $.
The portion of the characteristic equation associated with each eigenvalue $ \lambda_{K,i}$,  i.e.,  Eq.~\eqref{stability_characteristic_eq_new_0},  can be 
rewritten as 
\begin{align}
    \tau ( s_i -\lambda_{i}  ) e^{\tau ( s_i -\lambda_{i}  )}  & =      \tau \lambda_{d,i} e^{- \lambda_{i} \tau} , 
\label{stability_characteristic_eq_new}
\end{align} 
which can be solved numerically using the Lambert W function~\cite{Corless1996}  
$W(H) e^{W(H)}   =     H$
as 
\begin{align}
s_{i,\hat{k}}   & =    \lambda_{i}  +  \frac{1}{\tau}W_{\hat{k}}\left( \tau \lambda_{d,i} e^{- \lambda_{i} \tau} \right)
\label{Lambert_W_function_roots} 
\end{align}
for the $\hat{k}^{th}$ branch of the Lambert W function. 
Solutions to the nonhomogeneous  DDE~\eqref{system_DR} with nonzero source $z_s(t)$
can be specified using the roots in Eq.~\eqref{Lambert_W_function_roots}, especially since matrices $A$ and $A_d$ commute, e.g., see~\cite{Asl_Ulsoy_2003}. 
However, such numerical methods do not lead to a stability guarantee, which is addressed in the following subsection.

  \vspace{0.1in}
 \subsection{Condition for DSR stability}
  \vspace*{-0.1in}\noindent
A condition for stability of the DDE in Eq.~\eqref{system_DR} is developed  below, under the 
following assumption.

\begin{assumption}[Selection of controller]
\label{assumption_controller_parameter}
The DSR parameter $\beta >0$ is chosen to be sufficiently large, i.e., 
 \begin{align}
\beta  >     \max_{1 \le i \le n}     \frac{ 1 } {  {{\mathcal{R}}e} \left\{ {  \lambda_{K,i}  }\right\}   } > 0 , 
\label{Condition_Assumption_stability_filter_4}
\end{align}
where   $\left\{ \lambda_{K,i} \right\}_{i=1}^{n}$  are 
the  (potentially repeated) eigenvalues
of matrix $K$ with positive real parts.  \hfill  \qed
 \end{assumption}

\begin{Theorem}[Exponentially stability]
\label{theorem_Asymptotic_stability}
Under Assumptions \ref{assumption_digraph_properties} and \ref{assumption_controller_parameter}, 
the DDE system in Eq.~\eqref{system_DR} 
is exponentially stable, 
if 
\begin{align}
 \left| \beta \lambda_{K,i} -1  \right|   - \left( \beta  {{\mathcal{R}}e} \left\{ {  \lambda_{K,i}  }\right\}  -1     \right) 
 & <   \alpha  {\tau} {{\mathcal{R}}e}  \left\{ { \beta  \lambda_{K,i}  
  }\right\} 
 \label{eq_theorem_1_condition}
\end{align}
for all eigenvalues  $\left\{ \lambda_{K,i} \right\}_{i=1}^{n}$  
of matrix $K$.
\end{Theorem}

\noindent
{\textbf{Proof}}~
The proof aims to show that all roots  of the characteristic Eq.~\eqref{stability_characteristic_eq} have negative real parts (lie to the left of the imaginary axis of the complex plane) and then 
stability follows from Lemma~\ref{Lemma_DDE_eigenvalues}. 
The proof is through contradiction.  Assume that there is a root $s_i = a_i + j b_i$ with nonnegative real part
\begin{align}
{{\mathcal{R}}e\{s_i\}}  =  {{\mathcal{R}}e\{ a_i + j b_i \}}   = a_i  \ge   0   
\label{eq_stability_proof_0}
\end{align} 
 that satisfies the characteristic Eq.~\eqref{stability_characteristic_eq}, e.g., 
\begin{align}
 s_i + \alpha\beta \lambda_{K,i}  -\frac{ 1 }{\tau}\left[ 1  -\beta \lambda_{K,i} \right] \left(1 - e^{-s_i \tau} \right)   = 0   
\label{eq_stability_proof_1}
\end{align} 
for some eigenvalue 
$ \lambda_{K,i} $ of the pinned Laplacian $K$.
Then, from the nonnegative real part assumption $ {{\mathcal{R}}e} \left\{{s_i}\right\} \ge 0$  in Eq.~\eqref{eq_stability_proof_0} and from Eq.~\eqref{eq_stability_proof_1} 
\begin{align}
 {{\mathcal{R}}e}  \left\{ {- \alpha\beta \tau \lambda_{K,i}  
 +\left[ 1  -\beta \lambda_{K,i} \right] \left(1 - e^{-s_i \tau} \right)
 }\right\} \ge  0 .
   \label{eq_stability_proof_2}  
\end{align} 
Note that the set of points defined by  the second term $S_i(s)$,  
\begin{align}
S_i(s) ~ = \left[ 1  -\beta \lambda_{K,i} \right] \left(1 - e^{-s \tau} \right), 
 \label{Proof_thorem_Si}
\end{align} 
where $s$ has nonnegative real part,  is bounded by the circle found by evaluating 
$S_i(s)$ on the imaginary axis. The  circle is centered at $C_i = 1  -\beta \lambda_{K,i}$ 
and its radius ${\rho}_i$ is given by 
the magnitude of $C_i$, i.e., 
${\rho}_i = |C_i | =  | 1  -\beta \lambda_{K,i} | > 0$. 

\suppressfloats
\begin{figure}[!ht]
\begin{center}
\includegraphics[width=.45\columnwidth]{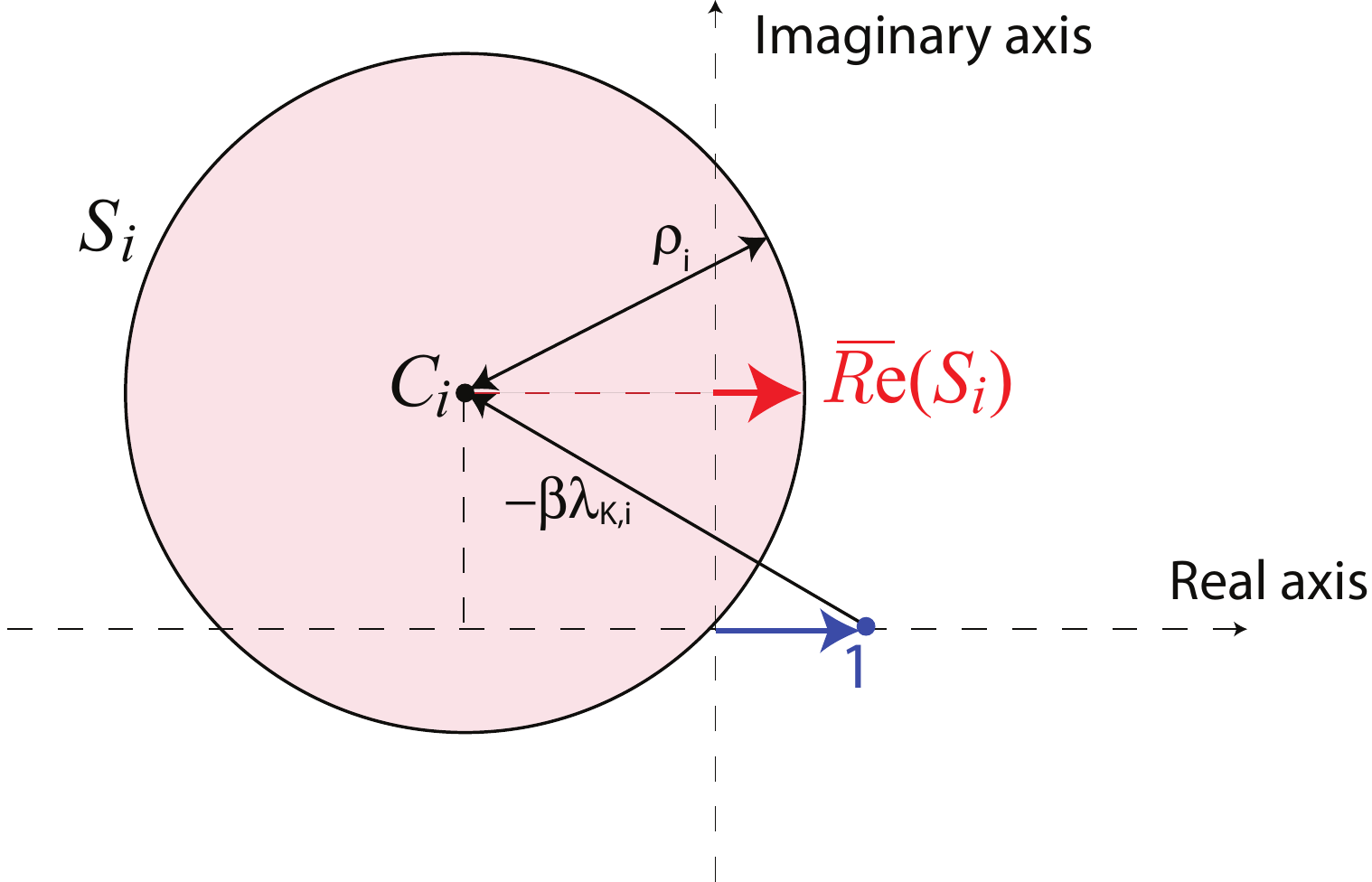}
\vspace{0.01in}
\caption{Set $S_i(s)$ in Eq.~\eqref{Proof_thorem_Si} when $s$ has nonnegative real part 
}
\label{fig_3_proof}
\end{center}
\end{figure}

Therefore, the maximum real part of $S_i(s)$ is 
 achieved on the imaginary axis, i.e., 
 \begin{align}
 \max_{ {{\mathcal{R}}e} \left\{{s}\right\} \ge 0 } 
{{\mathcal{R}}e} \left\{ S_i(s)   \right\}
 ~~ = 
  \max_{ {{\mathcal{R}}e} \left\{{s}\right\} = 0 } 
{{\mathcal{R}}e}  \left\{ S_i(s)   \right\}.
\end{align}  
Note that the  real part of the center $C_i$ is negative 
$$
 {{\mathcal{R}}e} \left\{ { C_i  }\right\} 
~ = 
 {{\mathcal{R}}e} \left\{ {  1- \beta \lambda_{K,i}  }\right\} ~= 1 - \beta  {{\mathcal{R}}e} \left\{ {  \lambda_{K,i}  }\right\} < 0 
$$
from Assumption~\ref{assumption_controller_parameter}. 
Therefore, the maximum 
real part of  $S_i(s)$  (for $s$ with nonnegative real part) is given by 
 \begin{align}
 \overline{  {{\mathcal{R}}e} } \left( S_i \right)~ & =
  \max_{ {{\mathcal{R}}e} \left\{{s}\right\} \ge 0 } 
{{\mathcal{R}}e}  \left\{ S_i(s)   \right\}
= ~   \left| {\rho}_i   \right|  - \left|  {{\mathcal{R}}e} \left\{{  C_i }\right\} \right|  \nonumber \\
& = \left| 1- \beta \lambda_{K,i}  \right|   - \left( \beta  {{\mathcal{R}}e} \left\{ {  \lambda_{K,i}  }\right\}  -1     \right) .
 \label{eq_theorem_proof_4} 
\end{align} 
From Eq.~\eqref{eq_stability_proof_2} and Eq.~\eqref{eq_theorem_proof_4}, the assumption that the real part of the 
root $s_i$ is nonnegative implies that 
\begin{align}
- \alpha\beta \tau{{\mathcal{R}}e}  \left\{ { \lambda_{K,i}  }\right\}
+  {{\mathcal{R}}e} \left\{ { S_i(s_i) }\right\} &  \ge  0 
   \label{eq_stability_proof_5}  
\end{align} 
or 
\begin{align}
{{\mathcal{R}}e} \left\{ { S_i(s_i) }\right\} &  \ge   \alpha\beta \tau{{\mathcal{R}}e}  \left\{ { \lambda_{K,i}  }\right\} 
   \label{eq_stability_proof_6} .
\end{align}
This requires the maximum possible value of $S_i(s)$ with positive real parts to be larger than the right hand side, i.e.,  
\begin{align}
 \max_{ {{\mathcal{R}}e} \left\{{s}\right\} \ge 0 } 
{{\mathcal{R}}e}   \left\{ S_i(s)   \right\} &  \ge   \alpha\beta \tau{{\mathcal{R}}e}  \left\{ { \lambda_{K,i}  }\right\} 
   \label{eq_stability_proof_7} 
\end{align}
or 
\begin{align}
\left| 1- \beta \lambda_{K,i}  \right|   - \left( \beta  {{\mathcal{R}}e} \left\{ {  \lambda_{K,i}  }\right\}  -1     \right) 
 &  \ge   \alpha\beta \tau{{\mathcal{R}}e}  \left\{ { \lambda_{K,i}  }\right\} , 
   \label{eq_stability_proof_7} 
\end{align}
which contradicts the condition in Eq.~\eqref{eq_theorem_1_condition}.
~~~~~~~~~ \hfill \qed \\

\vspace*{-0.1in}\noindent
 The stability condition in Theorem~\ref{theorem_Asymptotic_stability} 
 can be restated in terms of the known range of the eigenvalues 
  $ \lambda_{K,i} $ 
  of the pinned Laplacian $K$ .
  
 \begin{assumption}[Range of eigenvalues]
\label{assumption_range_eigenvalues}
\hfill The eigenvalues
$  \lambda_{K,i} =  m_{i}e^{j \phi_{i}}, $
 lie in the range specified by 
 \begin{align}
  0  & < \underline{m}   \le m_i    \le \overline{m},     \quad  ~ 
| \phi_i |    \le \overline{\phi}      < \frac{\pi}{2}, 
\label{Condition_Assumption_range_of_lambda}
\end{align}
where   the zero lower bound on the magnitude $m_i$ and the upper bound $\frac{\pi}{2} $ on the phase $\phi_i$   
arise since the eigenvalues $\lambda_{K,i}$  have positive real parts, as in Eq.~\eqref{eq_real_part_K}.   
\hfill  \qed
 \end{assumption}

\begin{corollary}[Range-based stability]
\label{Cor_DDE_range_based_stability}~

\vspace{-0.1in}
\noindent  
Under Assumptions~\ref{assumption_controller_parameter} and~\ref{assumption_range_eigenvalues}, the stability condition in 
Eq.~\eqref{eq_theorem_1_condition} of Theorem~\ref{theorem_Asymptotic_stability}  is met if 
\begin{align}
\frac{\overline{\rho} ~+1}{\overline{\rho}  \cos{\overline{\psi}}~ +1 }  -1  < \alpha  {\tau} 
 \label{eq_corollary_11_condition_1}
\end{align}
where, with $\beta { \underline{m} \cos{\overline{\phi} }  }   >  1 $ to satisfy Assumption~\ref{assumption_controller_parameter}, 
\begin{equation}
\begin{aligned}
\overline{\rho} &  =   \sqrt{  \left(  \beta \overline{m} \sin{\overline{\phi} } \right)^2 
+ \left(  \beta \overline{m} \cos{\overline{\phi} } -1 \right)^2 
}
 \\
 \overline{\psi}  & =  \tan^{-1}\left(  \frac{ \beta \underline{m} \sin{\overline{\phi} }}{ 
 \beta \underline{m} \cos{\overline{\phi} } -1} \right) .
   \label{eq_corollary_11_condition_2} 
 \end{aligned}
\end{equation}

\end{corollary}

\begin{figure}[!ht]
\begin{center}
\includegraphics[width=.6\columnwidth]{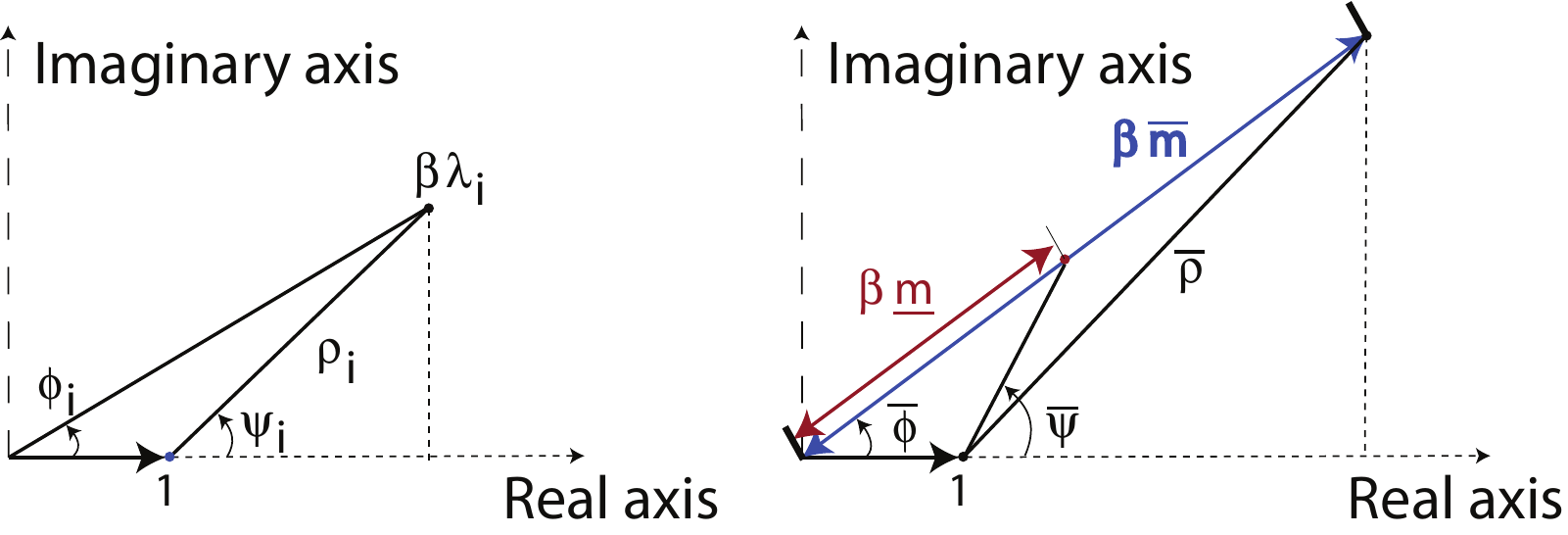}
\vspace{0.01in}
\caption{(Left) Stability condition terms in Eq.~\eqref{eq_corollary_proof_1} for a specific eigenvalue $\lambda_i$. (Right) 
Terms in Eq.~\eqref{eq_corollary_11_condition_1} based on  range of eigenvalue $\lambda_i$ in 
Assumption~\ref{assumption_range_eigenvalues}. 
}
\label{fig_corollary_proof}
\end{center}
\end{figure}

\noindent
{\textbf{Proof}}~
Note that the real part of $\beta \lambda_{K,i}  $ is greater than one from Eq.~\eqref{Condition_Assumption_stability_filter_4}.
Let $ |  \beta  \lambda_{K,i}  -1 | = {\rho}_i$, 
and $\beta  {{\mathcal{R}}e} \left\{ {  \lambda_{K,i}  }\right\}  -1    =  {\rho}_i \cos{ \psi_i} $
as in Fig.~\ref{fig_corollary_proof}. 
 \suppressfloats
Therefore, the stability condition in 
Eq.~\eqref{eq_theorem_1_condition} can be rewritten as 
\begin{align}
\frac{{\rho}_i  - {\rho}_i \cos{ \psi_i} }{ {\rho}_i \cos{ \psi_i} + 1}
~  
= 
\frac{{\rho}_i  +1 }{ {\rho}_i \cos{ \psi_i} + 1} -1
~
 & <   \alpha  {\tau} , 
 \label{eq_corollary_proof_1}
\end{align}
where the left hand side (lhs) of the inequality is a monotonic (nondecreasing) function of each variable 
$\rho_i$ and $ \psi_i$, independent of the other variable, over the entire interval ${\rho}_i >0 $ 
and  $0 \le  \psi_i < \pi/2$, 
since 

\vspace{-0.2in}
{\small{
\begin{equation}
\begin{aligned}
\frac{ \partial}{ \partial   {\rho}_i } 
\left( \frac{{\rho}_i  +1 }{ {\rho}_i \cos{ \psi_i} + 1}   \right) 
& = 
 \frac{ 1 - \cos{ \psi_i} }{\left(  {\rho}_i \cos{ \psi_i} + 1   \right)^2 }  ~~ \ge 0, \\
 \frac{ \partial}{ \partial  {\psi}_i } 
\left( \frac{{\rho}_i  +1 }{ {\rho}_i \cos{ \psi_i} + 1}   \right) 
& = 
 \frac{   ({\rho}_i  +1){\rho}_i \sin{ \psi_i}  }{\left(  {\rho}_i \cos{ \psi_i} + 1   \right)^2 } ~~ \ge 0 .
\end{aligned}
 \label{eq_corollary_proof_1_2}
\end{equation}
}}

\vspace{-0.2in}
\noindent
Due to symmetry, only the top portion of the right-half plane is 
considered, i.e., $0 \le \phi_i  $, $ 0 \le \psi_i   < \pi/2$.
Therefore, the lhs of Eq.~\eqref{eq_corollary_proof_1} is maximized with 
the largest selection of ${\rho}_i$ and $\psi_i$.
The largest angle $\psi_i$ is 
$\overline{\psi}$ as in Eq.~\eqref{eq_corollary_11_condition_2} 
corresponding to the  smallest magnitude $\underline{m}$ and largest phase $\overline{\phi}$
in the range specified by Eq.~\eqref{Condition_Assumption_range_of_lambda} 
since the derivative of 
\begin{align}
\tan{\psi_i} =  \beta m_i \sin{\phi_i}/ \left( \beta m_i \cos{\phi_i} -1 \right) 
\end{align}
is nonpositive with respect to the magnitude $m_i$ and nonnegative with respect to  phase  $0 \le \phi_i < \overline{\phi}  $ 
of the eigenvalue  $\lambda_{K,i}$ satisfying Eq.~\eqref{Condition_Assumption_range_of_lambda}, 
\begin{equation}
\begin{aligned}
\frac{ \partial \tan{\psi_i}  }{ \partial  m_i}   
& = 
\frac{ -\beta \sin{\phi_i}   }{ 
\left( {  \beta {m_i} \cos{\phi_i}  -1} \right)^2 } 
 ~~ \le  0 , 
\\
\frac{ \partial \tan{\psi_i}  }{ \partial  \phi_i}   
& = 
\frac{  \beta m_i ( \beta m_i -\cos{\phi_i})   }{ 
\left( {  \beta {m_i} \cos{\phi_i}  -1} \right)^2 } 
~~ \ge 0 
\end{aligned}
 \label{eq_corollary_proof_1_4}
\end{equation}
as $ \beta m_i \ge  \beta m_i \cos{\phi_i} > 1 \ge  \cos{\phi_i} $ 
from Eq.~\eqref{Condition_Assumption_stability_filter_4}.
Similarly, the largest magnitude  ${\rho}_i$ is $\overline{\rho} $ is 
obtained by choosing 
the largest magnitude $\overline{m}$ and largest phase $\overline{\phi}$ for the 
eigenvalues $\lambda_{K,i}$ satisfying Eq.~\eqref{Condition_Assumption_range_of_lambda}, as in Eq.~\eqref{eq_corollary_11_condition_2}. This is because the derivative of 
the square of the magnitude 
 \begin{equation}
\begin{aligned}
  {\rho}_i^2 & =  (\beta m_i \sin{( \phi_i)})^2 +  (\beta m_i \cos{( \phi_i)} -1 )^2   \\
  & =  \beta^2 m_i^2 -2 \beta m_i \cos{( \phi_i)} + 1  
\end{aligned} 
\end{equation}
is a nondecreasing  function of magnitude $ m_i>0$ and phase $\phi_i$,   with $  0< \phi_i \le \pi/2$, since 
 \begin{equation}
\begin{aligned}
\frac{ \partial {\rho_i}^2 }{ \partial  m_i}   
& = 
 2 \beta^2  m_i -2 \beta  \cos{( \phi_i)}    = 2 \beta \left( \beta m_i  -  \cos{( \phi_i)}     \right) 
\ge  0 
\\
\frac{ \partial {\rho_i}^2 }{ \partial  \phi_i }   
& = 
 2 \beta m_i \sin{( \phi_i)}  ~~ \ge 0.  
\end{aligned}
 \label{eq_corollary_proof_1_3}
\end{equation}
The theorem follows since the lhs of Eq.~\eqref{eq_corollary_11_condition_1} is an upper bound for 
the lhs of Eq.~\eqref{eq_corollary_proof_1}  \hfill \qed

  \vspace{-0.1in}\subsection{Topological ordering and rapid  cohesive transition}
   \vspace*{-0.1in}\noindent
 Arbitrarily-fast cohesive transition can be achieved (within constraints, such as actuator bandwidth) if the  eigenvalues of the pinned Laplacian $K$ are real, as shown below.   
Additionally, this section connects the stability analysis of the proposed DSR approach in Theorem~\ref{theorem_Asymptotic_stability}  with prior methods for  stability of DDEs. 
  
With real eigenvalues $\lambda_{K,i} $, the only condition for stability is that the parameter $\beta$ be sufficiently large, as in Assumption~\ref{assumption_controller_parameter}. This is stated formally below.
\begin{corollary}[Response rate with real eigenvalues]
\label{Cor_Symmetry_Rate_information_change}
Let the eigenvalues  $\lambda_{K,i} $  
of the pinned Laplacian $K$  be real, i.e., 
${{\mathcal{R}}e} \left\{ {  \lambda_{K,i}  }\right\}  =  \lambda_{K,i}$
for all $ 1 \le i \le n$ 
and  let the parameter $\beta$ be sufficiently large as in Eq.~\eqref{Condition_Assumption_stability_filter_4}. 
Then, the system with DSR
is stable for 
 any  positive choice of the cohesive  response rate $\alpha>0$ and the time delay 
 $\tau>0$.
\end{corollary}

\noindent
{\textbf{Proof}}~
With real eigenvalues $\lambda_{K,i}$, from Assumption~\ref{assumption_controller_parameter}, 
$$ \left|  \beta \lambda_{K,i} -1 \right|  = 
\beta \lambda_{K,i} -1, \qquad 
\left( \beta  {{\mathcal{R}}e} \left\{ {  \lambda_{K,i}  }\right\}  -1     \right)  
= \beta \lambda_{K,i} -1. 
$$ 
Therefore, the condition in Eq.~\eqref{eq_theorem_1_condition} of 
Theorem~\ref{theorem_Asymptotic_stability} becomes 
$0 <   \alpha  {\tau}  \beta  \lambda_{K,i}$, 
which is satisfied for 
any positive response-rate parameter $\alpha$ and delay $\tau$
since $\beta > 0$ and  $\lambda_{K,i} > 0$. 
 \hfill \qed

\begin{rem} [Undirected graph]
\label{Rem_Braytons_theorem}
Using Theorem~1 in~\cite{Brayton_67}, it can be shown that  the DDE   in Eq.~\eqref{system_DR} is stable
when the graph ${\mathcal{G}}\!\setminus\!s$  associated with the pinned Laplacian $K$  
is undirected, provided  $ -A \pm A_d$ is positive definite.
The positive definiteness of   $ -A \pm A_d$  
follows from  Corollary~\ref{Cor_Symmetry_Rate_information_change} when $\alpha > 0, \beta > 0$. 
Note that $ -A -A_d = \alpha \beta  K $  is positive definite since $K$ is positive definite.
Moreover,  $ -A +A_d = \alpha \beta  K - \frac{2}{\tau}\left[ I -  \beta K    \right] $ is positive definite since it is symmetric 
and its eigenvalues $\lambda_{AAd}$ are positive, i.e., 
$ \lambda_{AAd} = \alpha\beta  \lambda_{K,i}   -2\frac{1}{\tau}\left[ 1  - \beta \lambda_{K,i}  \right] > 0  $  
since $ \lambda_{K,i}   > 0 $ are the positive eigenvalues of the symmetric pinned Laplacian $K$ and 
$ \lambda_{K,i} \beta  > 1$ from Assumption~\ref{assumption_controller_parameter}.
\end{rem}

 \vspace{0.05in}
\begin{rem} [Necessary and sufficient conditions]
\label{Rem_Hayes_theorem}
The conditions of Corollary~\ref{Cor_Symmetry_Rate_information_change} meet 
the following necessary and sufficient conditions for stability developed in~\cite{Hayes_50} (when the parameters of the characteristic equation 
$ \lambda_{i},  \lambda_{d,i} $ are real),  i.e., 
\begin{align}
\lambda_{i}   < 1,   \quad  {\mbox{and}} \quad  \lambda_{i}    < -  \lambda_{d,i} < (V_\lambda^2 +  \lambda_{i}^2)^{\frac{1}{2}}, 
\label{stability_hayes}
\end{align} 
where $V_\lambda$ is the root of $  V_\lambda \cot{V_\lambda} =  \lambda_{i}   $ such that $0 <  V_\lambda  < \pi $. 
Note that $\lambda_{i}  $ from Eq.~\eqref{system_eigenvalues_DR_matrices} becomes 
$$
\lambda_{i}  =  - \alpha\beta \lambda_{K,i} +\frac{1}{\tau}\left[ 1  - \beta \lambda_{K,i} \right]  , 
$$
which is less than zero since $ \beta \lambda_{K,i} > 1$ and each term in $ \alpha\beta \lambda_{K,i}$ is positive. Moreover, $ \lambda_{i}    < -  \lambda_{d,i} $ from the 
definition in Eq.~\eqref{system_eigenvalues_DR_matrices} since the terms $\alpha\beta \lambda_{K,i} > 0 $,  and 
$-  \lambda_{d,i} < (V_\lambda^2 +  \lambda_{i}^2)^{\frac{1}{2}}$ since $|\lambda_{i} | > |\lambda_{d,i} | $ and $V_\lambda > 0 $.
 \hfill \qed
\end{rem}

When the pinned Laplacian $K$  associated with the graph ${\mathcal{G}}\!\setminus\!s$  of the  non-source agents is undirected, the pinned Laplacian $K$ 
is real symmetric  and therefore its eigenvalues are real. However, the eigenvalues of the pinned Laplacian $K$ can be real with directed 
graphs such as topologically-ordered subgraphs.

\begin{rem}[Topologically ordered graphs]
\label{rem_Toplogically_ordered_graphs}
For  acyclic directed graphs (or topologically ordered graphs),  the  pinned Laplacian $K$ 
will be lower-diagonal and hence have real eigenvalues. 
In an acyclic graph there is a 
topological ordering of the nodes ${\mathcal{V}}$ and every graph edge  ${\mathcal{E}}$ goes from a node that is earlier in the ordering to a node that is later in the ordering, i.e., all the neighbors
$N_i$ of a node $i$ are earlier in the ordering. This leads to a pinned Laplacian $K$ which is lower diagonal and real and hence, with real eigenvalues. 
 \hfill \qed
\end{rem}

\begin{rem}[Topologically-ordered sub-graphs]
\label{rem_Toplogically_ordered_sub_graphs}
The eigenvalues of the pinned Laplacian $K$ are real when the matrix $K$ is associated with a 
set of subgraphs ${\mathcal{G}}_i$ that are distinct (i.e., without shared nodes) where each subgraph is either symmetric or acyclic (topologically ordered)  
with an additional topological ordering of the subgraphs ${\mathcal{G}}_i$  such that all  
graph edges in ${\mathcal{G}}\!\setminus\!s$   ends in one of the subgraphs, say ${\mathcal{G}}_i$  and starts: 
(a)~either in the same ending subgraph ${\mathcal{G}}_i$; or (b) in a subgraph that is earlier than the ending subgraph ${\mathcal{G}}_i$ in the subgraph ordering. 
Such topological ordering of the subgraphs  ensures that the pinned Laplacian $K$,  associated with the graph  ${\mathcal{G}}\!\setminus\!s = \bigcup {\mathcal{G}}_i $   
is lower block-diagonal with symmetric matrices $K_i$ in each diagonal block. Then, the  eigenvalues of the 
pinned Laplacian $K$ are real because they are the same as the eigenvalues of the real-valued matrices $K_i$, each of which is either diagonal or symmetric. 
 \hfill \qed
\end{rem}

\vspace{-0.1in}
\begin{rem}[Impact of noise]
\label{rem_impact_of_noise_2}
Although stability is not impacted by the noise, the cohesion performance can deteriorate in the presence of substantial noise. 
If noise of size $N_v$ is present in the estimation of $v$ in Eq.~\eqref{system_non_source_i_2_mod}, then it leads to a noise of order 
$\frac{N_v}{\tau}$ due to the approximated derivative in Eq.~\eqref{system_non_source_i_mod}. 
Thus,  the time delay $\tau$ needs to be sufficiently large to reduce the noise effect 
on the dynamics, which in turn increases the achievable settling time $T_s$
as in Eq.~\eqref{eq_tau_Ts_link}. Alternatively, the noise can be filtered as shown in the following subsection. 
\end{rem}

 \subsection{DSR with higher-order dynamics}
  \vspace*{-0.1in}\noindent
The DSR approach can be extended 
to enable cohesive tracking 
when the agents have
higher-order dynamics, and the DSR update can be filtered to reduce noise effects, as shown below.

 \subsubsection{Agent's higher-order dynamics}
  \vspace*{-0.1in}\noindent
Let the dynamics of an individual 
agent $i$ be given by a minimum-phase system in the output-tracking form (through appropriate 
input and state transformations, e.g., see~\cite{isidori}) as 
\begin{equation}
\begin{aligned}
 {z}_i^{(r)} (t)  &  = u_i(t), \\
\frac{d}{dt} {\eta}_i (t)   & = A_{\eta,i}  {\eta}_i(t) + A_{z,i} Z_i(t),  
\label{system_hod}
\end{aligned}
\end{equation}
where $r$ is the relative degree (i.e., the  difference between the number of poles and the number of zeros), 
the bracketed superscript denotes the time derivative, e.g., 
$z_i^{(r)}$ represents the $r^{th}$ time derivative of $z_i$, and 
$Z_i$ represents the agent output  $z_i$ and its time derivatives $z_i^{(k)}$,  $1 \le k \le r-1$. 
The internal dynamics  represented by $\eta_i$  is stable, i.e., $A_{\eta,i} $ is Hurwtiz,  
since the system is minimum phase. 
Note that the stability of the internal dynamics is independent of the selection of the control input $u_i$. 
 
\vspace{-0.1in}
\begin{rem}[Heterogeneous agents]
\label{rem_different dynamics}
The internal dynamics ${\eta}_i $ in Eq.~\eqref{system_hod} can be different and can be nonlinear, 
provided the dynamics remain close to the stable origin of the internal dynamics.  
In this sense, the approach is applicable to heterogeneous agents. 
\end{rem}

\begin{assumption}[Relative degree]
\label{assumption_Relative_degree}
All agents have  minimum-phase dynamics and  the same well-defined relative degree  $1 \le r \le n$. 
\hfill  \qed
 \end{assumption}

 \subsubsection{Ideal cohesive higher-order dynamics}
 \vspace*{-0.1in}
 \noindent
 If each non-source agent can have instantaneous access to the source $z_s$ (which is sufficiently smooth), 
 then the input $u_i(t)$ can be selected such that the output $z_i$ dynamics is given by, as in Eq.~\eqref{system_limit_single}, 
 \begin{align}
 {z}_i^{(r)} (t)  &  =  {z}_s^{(r)} (t)   -\sum_{k=0}^{r-1}  \hat\alpha_{k} \left( {z}_i^{(k)} (t)  -  {z}_s^{(k)} (t)  \right).  
\label{system_limit_single_hod}
\end{align}
With the same initial condition,  the response of all the agents would track the desired output $z_s$
in a cohesive manner. The output tracking is stable  since the characteristic equation ${\mathcal{P}}(s) =0$, 
 \begin{align}
 {\mathcal{P}}(s)  =   \sum_{k=0}^{r} \hat\alpha_{k} s^{k}   = (s + \alpha)^r
 \label{system_polynomial_hod}
 \end{align}
 has stable roots provided $\alpha > 0$. Note that the leading coefficient  is one, i.e., $ \hat\alpha_{r} =1$, 
 and the  constant $\alpha$ can be varied to adjust the 
overall speed of the response of each agent. 
In a vector form, this ideal cohesive  dynamics can be written as 
\begin{align}
 {Z}^{(r)} (t)  &  = 
 -\sum_{k=0}^{r-1}  \hat\alpha_{k}  {Z}^{(k)} (t) 
 + 
 {\textbf{1}}_n 
  \sum_{k=0}^{r}  \hat\alpha_{k}  {z}_s^{(k)} (t) .
\label{system_limit_hod}
\end{align}
\noindent
Multiplying both sides of Eq.~\eqref{system_limit_hod} by $\beta K$  (where $\beta > 0$) and using Eq.~\eqref{eq_KinvtimesB_2} to replace $K {\textbf{1}}_n$ results in 
\begin{align}
 \beta K  {Z}^{(r)} (t)  &  =
  -\beta K \sum_{k=0}^{r-1}  \hat\alpha_{k}  {Z}^{(k)} (t) 
 + 
 \beta B z_s^*(t), 
 \label{system_imit_mod1_hod_temp}
\end{align}
where 
$
z_s^*(t) =   \sum_{k=0}^{r}  \hat\alpha_{k}  {z}_s^{(k)} (t)
$.  
By adding ${Z}^{(r)} $ on both sides of Eq.~\eqref{system_imit_mod1_hod_temp},  the ideal cohesive  dynamics can be rewritten  as 
\begin{align}
{Z}^{(r)} (t)   &  =   -\beta K \sum_{k=0}^{r-1}  \hat\alpha_{k}  {Z}^{(k)} (t)   
 + \beta    B  z_s^* (t)  + \left[ I - \beta K \right] {Z}^{(r)}  (t). 
\label{system_R_hod}
\end{align}

 \vspace{-0.1in}
 \subsubsection{DSR-based implementation}
  \vspace*{-0.1in}\noindent
Approximating the derivative ${Z}^{(r)}$ on the right hand side of Eq.~\eqref{system_R_hod} in terms of delayed versions of 
the system state, as in Eq.~\eqref{system_DR_delayed_derivative}, 
\begin{align} 
{Z}^{(r)}(s)  & \approx  \hat{Z}^{(r)}(s,\tau) = \left[ f(s) \frac{ 1 - e^{-\tau s} }{\tau} \right]^r Z(s) = {\mathcal{F}}(s) Z(s), 
\label{system_DR_delayed_derivative_approx_hod}
\end{align}
where $f(s)$ is a low-pass filter, 
yields  the DSR  approach for networks with higher-order dynamics. Replacing ${Z}^{(r)}(t)$ with 
$\hat{Z}^{(r)}(t,\tau)$, the Laplace inverse of $\hat{Z}^{(r)}(s,\tau)$, Eq.~\eqref{system_R_hod} becomes 
\begin{align}
{Z}^{(r)} (t)   &  =   -\beta K \sum_{k=0}^{r-1}  \hat\alpha_{k}  {Z}^{(k)} (t)   
 + \beta    B  z_s^* (t)  
 + \left[ I - \beta K \right] \hat{Z}^{(r)}(t,\tau). 
\label{system_DR_tracking_hod}
\end{align}
As in the first-order case, 
the DSR-based extension for the case when agents have higher-order dynamics only uses local information 
from the neighbors and does not require network changes.

 \subsection{Stability of DSR approach}
  \vspace*{-0.1in}\noindent
Stability of the DSR approach in Eq.~\eqref{system_DR_tracking_hod} depends on the 
roots of the characteristic equation 
\begin{equation}
\det \left|  s^r I +\beta K \sum_{k=0}^{r-1}  \hat\alpha_{k}  s^{k}      
-\left( I - \beta K \right) {\mathcal{F}}(s)  
     \right|  = 0.
\label{characteristic_eq_Miranker_hod}
\end{equation}
In particular, using arguments similar to those in Lemma~\ref{Lemma_DDE_eigenvalues}, 
the DSR-based approach is stable if and only if  
the roots 
$s_i$ of 
\begin{align}
 s^r  +\beta \lambda_{K,i}  \sum_{k=0}^{r-1}  \hat\alpha_{k}  s^{k}      
-\left( 1- \beta \lambda_{K,i}  \right) {\mathcal{F}}(s)  
     = 0
\label{stability_characteristic_eq_new_0_hod}
\end{align}
have negative real part where 
$\left\{ \lambda_{K,i} \right\}_{i=1}^{n}$  are eigenvalues of the pinned Laplacian $K$. 
Equation~\eqref{stability_characteristic_eq_new_0_hod} can be rewritten using Eq.~\eqref{system_polynomial_hod} 
as  
\begin{align}
(1 -\beta \lambda_{K,i} )   s^r   +\beta \lambda_{K,i} {\mathcal{P}}(s)      
& -\left( 1- \beta \lambda_{K,i}  \right) 
 {\mathcal{F}}(s) 
     = 0
\nonumber  
\end{align}
or, since $ 1- \beta \lambda_{K,i} \ne 0 $ under 
Assumption~\ref{assumption_controller_parameter}, 
\begin{align}
   s^r  & + \frac{\beta \lambda_{K,i}}{(1 -\beta \lambda_{K,i} )}  {\mathcal{P}}(s)     = {\mathcal{F}}(s) .
\label{proof_hod_2}
\end{align}

\begin{Theorem}[Exponentially stability of extended DSR]
\label{theorem_Asymptotic_stability_extended}
Under Assumptions~\ref{assumption_digraph_properties}-\ref{assumption_Relative_degree}, 
the  origin $Z(t) = 0$ of the DDE system in Eq.~\eqref{system_DR_tracking_hod} 
is exponentially stable, 
if 
\begin{equation}
\begin{aligned}
\sup_{{\mathcal{R}}e(s) \ge 0}   
\left| f(s) \left( 1 - e^{-\tau s} \right) \right|   & <   \left(  \epsilon_\lambda \right)^{\frac{1}{r}}  \alpha \tau , 
\end{aligned}
\label{eq_theorem_2_conditions}
\end{equation}
where 
\begin{align}
 \epsilon_\lambda  & = 
 \frac{\beta {\overline{m}} }{\sqrt{\beta^2 \overline{m}^2 - 2 \beta \overline{m} \cos{\overline\phi}  +1}}    
 \quad ~{\mbox{if}}~ r > 1 
 \label{eq_theorem_2_epslambda_condition_1}
 \\
 & = 
 \frac{\beta {\underline{m}} \cos{\overline\phi}  }{\sqrt{\beta^2 \overline{m}^2 - 2 \beta \overline{m} \cos{\overline\phi}  +1}}    
 \quad ~{\mbox{if}}~ r = 1
  \label{eq_theorem_2_epslambda_condition_2}
\end{align}
\end{Theorem}

\noindent
{\textbf{Proof}}~
This is shown by contradiction. Assume that $s_i  = a_i + j b_i $ is a root of Eq.~\eqref{proof_hod_2} 
with nonnegative real part $a_i \ge 0$. Then, with $s=s_i$, it is  shown below that, under the theorem's condition,  
that the smallest magnitude of 
the right hand side (rhs) of Eq.~\eqref{proof_hod_2}  is greater than the largest magnitude of
the left hand side (lhs), which contradicts the assumption that $s_i$ is a root of Eq.~\eqref{proof_hod_2}. 

\vspace{0.1in}
\noindent 
First, the case when the relative degree $r \ge 2$ is considered. 
Consider the magnitude of the factor multiplying $ {\mathcal{P}}(s) $ in Eq.~\eqref{proof_hod_2}.
Note that, for any eigenvalue 
$\lambda_{K,i}$ of the pinned Laplacian $K$ satisfying
Assumption~\ref{assumption_controller_parameter}, $ 0<  \beta m_i \cos{\phi_i} -1  < \beta m_i \cos{\phi_i}, $ and 
therefore 
\begin{equation}
\begin{aligned}
\left| \frac{\beta \lambda_{K,i}}{(1- \beta \lambda_{K,i} )} \right| & = 
\frac{   \sqrt{ (\beta m_i \cos{\phi_i}  )^2 + (\beta m_i \sin{\phi_i} )^2 }   }{
  \sqrt{ (\beta m_i \cos{\phi_i} -1 )^2 + (\beta m_i \sin{\phi_i} )^2 }   }  > 1  
\end{aligned}
\label{proof_Th2_1}
\end{equation}
for all magnitude $m_i$ and phase $\phi_i$ satisfying Assumption~\ref{assumption_range_eigenvalues}. 
The square of lhs of the inequality is 
monotonic non-increasing with both magnitude $m_i>0$ and phase $0\le \phi_i < \overline\phi$ 
since  
\begin{equation}
\begin{aligned}
\frac{\partial}{ \partial  m_i}
\left|
\frac{\beta \lambda_{K,i}}{(1- \beta \lambda_{K,i} )} \right|^2  
& = 
\frac{ -2\beta^2 m_i \left( \beta m_i \cos{\phi_i}  -1 \right) }{ 
\left| 1- \beta \lambda_{K,i} \right|^4 } 
 \le   0, 
\\
\frac{\partial}{ \partial  \phi_i}
\left|
\frac{\beta \lambda_{K,i}}{(1- \beta \lambda_{K,i} )} \right|^2  
& = 
\frac{-2 \beta^3 m_i^3  \sin{\phi_i}   }{ 
\left| 1- \beta \lambda_{K,i} \right|^4 }  
 \le 0 .
\end{aligned}
 \label{proof_Th2_2}
\end{equation}
Therefore, from Eqs.~\eqref{Condition_Assumption_range_of_lambda}, \eqref{proof_Th2_1} and \eqref{proof_Th2_2}, 
 \begin{equation}
\begin{aligned}
\left| \frac{\beta \lambda_{K,i}}{(1- \beta \lambda_{K,i} )} \right| 
\ge 
\frac{\beta {\overline{m}} }{\sqrt{\beta^2 \overline{m}^2 - 2 \beta \overline{m} \cos{\overline\phi}  +1}}    
= \epsilon_\lambda   > 1 , 
\end{aligned}
\label{proof_Th2_3}
\end{equation}
and
\begin{equation}
\begin{aligned}
\left|    \frac{\beta \lambda_{K,i}}{(1- \beta \lambda_{K,i} )}   {\mathcal{P}}(s_i)  \right|
=  \left|    \frac{\beta \lambda_{K,i}}{(1- \beta \lambda_{K,i} )}  \right| \left|    {\mathcal{P}}(s_i)  \right|
 & \ge 
\epsilon_\lambda  \left|  {\mathcal{P}}(s_i)  \right| .
\end{aligned}
\label{proof_Th2_1_22}
\end{equation}

Moreover, for any root $s_i  = a_i + j b_i $  with nonnegative real part $a_i\ge0$, 
since $\alpha>0$  in Eq.~\eqref{system_polynomial_hod} for 
stable cohesive tracking, 
\begin{equation}
\begin{aligned}
 | s_i +  \alpha |^{k}    
  =  \left[ \left(a_i +  \alpha \right)^2 + b_i^2 \right]^{k/2}   
  >   \left[ a_i ^2 + b_i^2 \right] ^{k/2}  =  | s_i |^{k} \ge 0 , 
\end{aligned}
\nonumber 
\end{equation}
for all integers $k\ge 0$. 
Therefore, 
\begin{equation}
\left|  {\mathcal{P}}(s_i)  \right| =  | s_i +  \alpha |^{r}   >  \left|  s_i \right|^r
\label{proof_Th2_1_3_2}
\end{equation}
and since  $ \epsilon_\lambda  > 1 $ from Eq.~\eqref{proof_Th2_3}, from Eq.~\eqref{proof_Th2_1_22} and 
\eqref{proof_Th2_1_3_2}, 
\begin{equation}
\begin{aligned}
\left|   s_i^r +  \frac{\beta \lambda_{K,i}}{(1- \beta \lambda_{K,i} )}   {\mathcal{P}}(s_i)  \right| & \ge 
\epsilon_\lambda  | s_i +  \alpha |^{r}  -  \left|s_i  \right|^r > 0.
\end{aligned}
\label{proof_Th2_1_3}
\end{equation}

To find the smallest possible magnitude of the lhs of Eq.~\eqref{proof_hod_2}, the 
difference in magnitudes of its two terms, i.e.,  the rhs  of Eq.~\eqref{proof_Th2_1_3}, is compared below. 
Note that the difference $H(s_i,k)$
\begin{equation}
\begin{aligned}
H(s_i,k)  =   \epsilon_\lambda  | s_i + \alpha|^k     - | s_i |^k    ~~> 0 
\end{aligned}
\label{proof_rep_pole_1}
\end{equation}
is positive 
for all $k\ge0$.  
Moreover, the  difference $H(s_i,r)$, between the magnitudes of the terms 
in the lhs of Eq.~\eqref{proof_hod_2}, 
is also a monotonic (non-decreasing) function 
with the components $a_i \ge 0$ and  $ b_i^2 \ge 0 $ of the root $s_i$, since
\begin{equation}
\begin{aligned}
\frac{ \partial H(s_i,r) }{ \partial  b_i^2 }   
& = 
 \frac{r  \epsilon_\lambda}{2} 
  \left| (a_i + \alpha)^2 +b_i^2 \right|^{\frac{r}{2}-1}
  -   \frac{r }{2}  \left| a_i ^2 +b_i^2 \right|^{\frac{r}{2}-1} \\
  = &  
 \frac{r}{2}  \left[  \epsilon_\lambda    | s_i + \alpha |^{r-2} -  | s_i  |^{r-2}  \right]  ~ =  \frac{r}{2}  H(s_i,r-2)  > 0, \\
\frac{ \partial H(s_i,r) }{ \partial  a_i }   
  &  = 
 r  \left[  \epsilon_\lambda   | s_i + \alpha |^{r-2} -  | s_i  |^{r-2}  \right] a_i    +
 r   \epsilon_\lambda   | s_i + \alpha |^{r-2}  \alpha  
\\ = &   r a_i  H(s_i,r-2)    +  r   \alpha \epsilon_\lambda   | s_i + \alpha |^{r-2}  
 \ge  
r   \epsilon_\lambda   \alpha^{r-1}  >0  
\end{aligned}
\nonumber
\end{equation}
when $r \ge 2$. 
Therefore, the magnitude difference $H(s_i,r)$, i.e., the smallest possible magnitude on the lhs of Eq.~\eqref{proof_hod_2}  
is bounded from below by (when $a_i = b_i =0$)
\begin{equation}
\begin{aligned}
H(s_i,r) & >  \epsilon_\lambda   \alpha^r .    
\end{aligned}
\label{proof_rep_pole_4}
\end{equation}
If this lower bound on the magnitude difference $H(s_i,r)$  of the lhs  is larger than the 
maximum magnitude of the rhs  ${\mathcal{F}}(s) $ in Eq.~\eqref{proof_hod_2}, then $s_i$ cannot be a root, i.e., 
a contradiction is obtained if 
\begin{equation}
\begin{aligned}
\epsilon_\lambda   \alpha^r   & >  
\sup_{{\mathcal{R}}e(s) \ge 0}   
 \left| f(s) \frac{ 1 - e^{-\tau s} }{\tau} \right|^r, 
\end{aligned}
\label{proof_rep_pole_5}
\end{equation}
which follows from the condition in Eq.~\eqref{eq_theorem_2_conditions}. 

Second, for the case when the relative degree is one, $r=1$, the magnitude 
of the lhs of Eq.~\eqref{proof_hod_2} is given by
%
%
%
\begin{equation}
\begin{aligned}
\left| s_i  + \frac{\beta \lambda_{K,i}}{(1 -\beta \lambda_{K,i} )}  {\mathcal{P}}(s_i)   \right|    
& 
 = 
\frac{ 
\left| (a + j b) + \alpha \beta \lambda_{K,i}   \right|    
}
{ \left| (1 -\beta \lambda_{K,i} )  \right| 
}
~~
\ge 
\frac{ 
 \beta {{\mathcal{R}}e}(\lambda_{K,i})    
}
{ \left| (1 -\beta \lambda_{K,i} )  \right|  
}  \alpha
\end{aligned}
\label{eq_thm2_proof_part_2}
\end{equation}
since $a\ge0$ and real part of $\lambda_{K,i}$ is positive. 
Note that 
the numerator ${{\mathcal{R}}e}(\lambda_{K,i}) $  is greater than or equal to the minimum possible value, 
$  \underline{m} \cos{\overline{\phi} } $,  
and the denominator $\left| (1 -\beta \lambda_{K,i} )  \right| $ 
is maximized by $\overline{\rho}$ from Eq.~\eqref{eq_corollary_11_condition_2}.
Therefore, the rhs of Eq.~\eqref{eq_thm2_proof_part_2} is bounded from below by $\epsilon_\lambda \alpha   ~> 0$. 
Again, if this minimum magnitude $\epsilon_\lambda\alpha $ of the lhs is larger than the rhs of Eq.~\eqref{proof_hod_2},
i.e., the condition in Eq.~\eqref{eq_theorem_2_conditions} is met, then $s_i$ with a nonnegative real part 
cannot be a root of Eq.~\eqref{proof_hod_2}. 
 \hfill \qed

\vspace{-0.1in}
\begin{rem}[Impact of filter]
\label{rem_filter_impact}
The filter $f(s)$ can be used to reduce the impact of noise with the delay-based approximation of the 
derivative in Eq.~\eqref{system_DR_delayed_derivative_approx_hod}.  
Note that the expression $\left( 1 - e^{-\tau s} \right)$ in 
 Condition~\eqref{eq_theorem_2_conditions} tends to zero when $s$ is small. 
 Therefore,  the use of an appropriate low pass filter   (e.g., sufficiently small $\Omega$ when 
 $f(s)= \Omega/(s+\Omega) $) can ensure that the lhs is small enough 
to satisfy Condition~\eqref{eq_theorem_2_conditions} for any delay $\tau$. However, aggressive filtering can 
reduce the effective bandwidth of the DSR-based cohesiveness. 
\end{rem}

\vspace{-0.1in}
\begin{rem}[Comparison of DSR stability conditions]
\label{rem_filter_impact}
For first-order agents, the stability condition Eq.~\eqref{eq_theorem_2_conditions} in Theorem~\ref{theorem_Asymptotic_stability_extended}, used to extend the DSR approach to higher-order agents (using magnitude-based arguments), is more conservative 
than the condition in Eq.~\eqref{eq_theorem_1_condition}  in Theorem~\ref{theorem_Asymptotic_stability} 
that uses real-component-based arguments for first-order agents. In particular, 
the lhs of in Eq.~\eqref{eq_theorem_1_condition}  of Theorem~\ref{theorem_Asymptotic_stability} is zero 
when the eigenvalues of the pinned Laplacian $K$ are real. In contrast, the lhs is nonzero, under the same 
conditions, in Eq.~\eqref{eq_theorem_2_conditions} of Theorem~\ref{theorem_Asymptotic_stability_extended}.
\end{rem}

\vspace{-0.1in}
\begin{rem}[Computational issues]
\label{rem_Computational_issues}
When the eigenvalues $\lambda_{K,i}$ of the 
pinned Laplacian $K$ are known, the lower bound $\epsilon_\lambda$ in 
Eqs.~\eqref{eq_theorem_2_epslambda_condition_1} and 
\eqref{eq_theorem_2_epslambda_condition_2}  can be replaced by 
${\overline{\epsilon_\lambda} } $ computed as 
\begin{align}
\epsilon_\lambda =  {\overline{\epsilon_\lambda} }  = 
\min_{ 1 \le i \le  n} \left| \frac{\beta \lambda_{K,i}}{(1- \beta \lambda_{K,i} )} \right|  
~\quad ~{\mbox{if}}~ r> 1 
\label{eq_hod_replace_overline_epsilon_lambda} \\
\epsilon_\lambda =  {\overline{\epsilon_\lambda} }  = 
\min_{ 1 \le i \le  n} \left| \frac{\beta {{\mathcal{R}}e}(\lambda_{K,i}) }{(1- \beta \lambda_{K,i} )} \right|  
~\quad ~{\mbox{if}}~ r= 1.
\label{eq_hod_replace_overline_epsilon_lambda_2} 
\end{align}
When the filter  has the form $ f(s) = \Omega/(s+\Omega)$ whose magnitude decreases when the real part of $s$ is positive and increasing, the lhs of the stability condition in Eq.~\eqref{eq_theorem_2_conditions} can be computed over the imaginary axis, i.e., 
\begin{equation}
\begin{aligned}
\sup_{{\mathcal{R}}e(s) \ge 0}   
\left| f(s) \left( 1 - e^{-\tau s} \right) \right|   = \sup_{{\mathcal{R}}e(s) = 0}   
\left| f(s) \left( 1 - e^{-\tau s} \right) \right|.
\end{aligned}
\label{eq_theorem_2_conditions_compute}
\end{equation}
\end{rem}

\vspace*{-0.2in}\noindent
\section{Simulation results and discussion }
\vspace*{-0.1in}\noindent
Simulation results are used to (i)~comparatively evaluate cohesion with and without DSR; and (ii)~to show the advantages of using DSR over attempting to 
increase the response speed for better cohesion.
While cohesion is expected to be better for smoother trajectories, the step response is used in the 
following  to quantify the cohesion, as described at the end of Section~2.

\vspace{-0.1in}\subsection{First-order example system}
\label{subsection_example_system}
\begin{figure}[!ht]
\begin{center}
\includegraphics[width=.5\columnwidth]{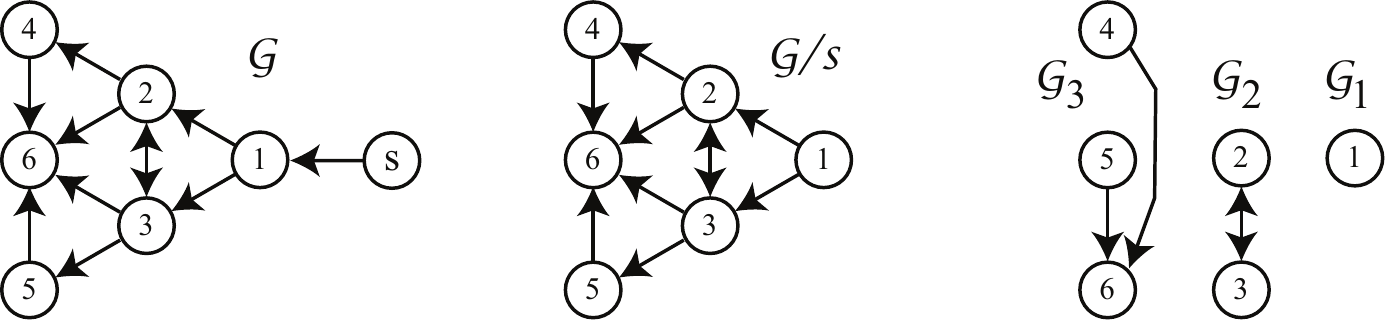}
\vspace{0.01in}
\caption{Network example used in simulations consisting of topologically ordered subgraphs  ${\mathcal{G}}_1 <  {\mathcal{G}}_2 <  {\mathcal{G}}_3$ of graph ${\mathcal{G}}\!\setminus\!s$ composed of undirected subgraphs ${\mathcal{G}}_1$, $ {\mathcal{G}}_2 $ and directed acyclic graph $ {\mathcal{G}}_3$ associated with node sets ${\mathcal{V}}_1 = \left\{ 1 \right\} , {\mathcal{V}}_2 = \left\{ 2, 3 \right\} $ and  ordered set ${\mathcal{V}}_3= \left\{ 4 <   5 <  6 \right\} $, 
respectively.   All edges in graph ${\mathcal{G}}\!\setminus\!s$  that ends in one of the subgraphs, say ${\mathcal{G}}_i$,  starts: 
(a)~either in the same subgraph ${\mathcal{G}}_i$; or (b) in a subgraph that is earlier than the subgraph ${\mathcal{G}}_i$ in the subgraph ordering. 
}
\label{fig_2_topological_ordering}
\end{center}
\end{figure}

  \vspace{-0.1in}\noindent
Consider an example system where the  graph ${\mathcal{G}}\!\setminus\!s$  is composed of  an ordered set of subgraphs ${\mathcal{G}}_1 <  {\mathcal{G}}_2 <  {\mathcal{G}}_3$  as shown in Fig.~\ref{fig_2_topological_ordering}, where ${\mathcal{G}}_1$ and $ {\mathcal{G}}_2 $ are undirected subgraphs  and  $ {\mathcal{G}}_3$  is an ordered acyclic graph associated with node sets ${\mathcal{V}}_1 = \left\{ 1 \right\} , {\mathcal{V}}_2 = \left\{ 2, 3 \right\} $ and  ordered set ${\mathcal{V}}_3= \left\{ 4 <   5 <  6 \right\} $, 
respectively. The associated  pinned Laplacian $K$,  where the weights $w_{ij}$ in Eq.~\eqref{eq_laplacian_defn} are either $0$ or $1$,  is

\begin{align}
\label{topological_ordering_example_K_2}
K & = 
\left[
\begin{array}{c|c|c} 
\begin{matrix}  1 \end{matrix} & 
\begin{matrix}    0    &  0  \end{matrix} & 
\begin{matrix}   0     & 0    &  0 \end{matrix} \\
\hline 
\begin{matrix}  -1  \\ -1   \end{matrix} & 
\begin{matrix}   2   &  -1  \\  -1    &   2 \end{matrix} & 
\begin{matrix}   0     & 0    &  0  \\  0      & 0    &   0  \end{matrix} \\
\hline
\begin{matrix}  0  \\ 0 \\ 0    \end{matrix} & 
\begin{matrix}  -1    &  0   \\  0   &  -1     \\  -1   &  -1  \end{matrix} & 
\begin{matrix}   1     & 0      &0 \\   0  &    1     & 0 \\   -1    & -1      & 4\end{matrix}
\end{array}
\right] 
= 
\left[
\begin{array}{c|c|c} 
K_1  & 
\begin{matrix}    0    &  0  \end{matrix} & 
\begin{matrix}   0     & 0    &  0 \end{matrix} \\
\hline 
\begin{matrix}  -1  \\ -1   \end{matrix} & 
K_2 & 
\begin{matrix}   0     & 0    &  0  \\  0      & 0    &   0  \end{matrix} \\
\hline
\begin{matrix}  0  \\ 0 \\ 0    \end{matrix} & 
\begin{matrix}  -1    &  0   \\  0   &  -1     \\  -1   &  -1  \end{matrix} & 
K_3
\end{array}
\right].
\nonumber
\end{align}

\suppressfloats

 \noindent
 The eigenvalues of the pinned Laplacian $K$ are then the 
 eigenvalue $1$ of $K_1$, eigenvalues $1, 3$ of $K_2$  and eigenvalues $1, 1, 4$ of $K_3$ for the example in Fig.~\ref{fig_2_topological_ordering}. 

\vspace{-0.05in}\subsection{Step response without DSR}
\begin{figure}[!ht]
\begin{center}
    \begin{tabular}{@{}cc@{}}
    \includegraphics[width=0.35\columnwidth]{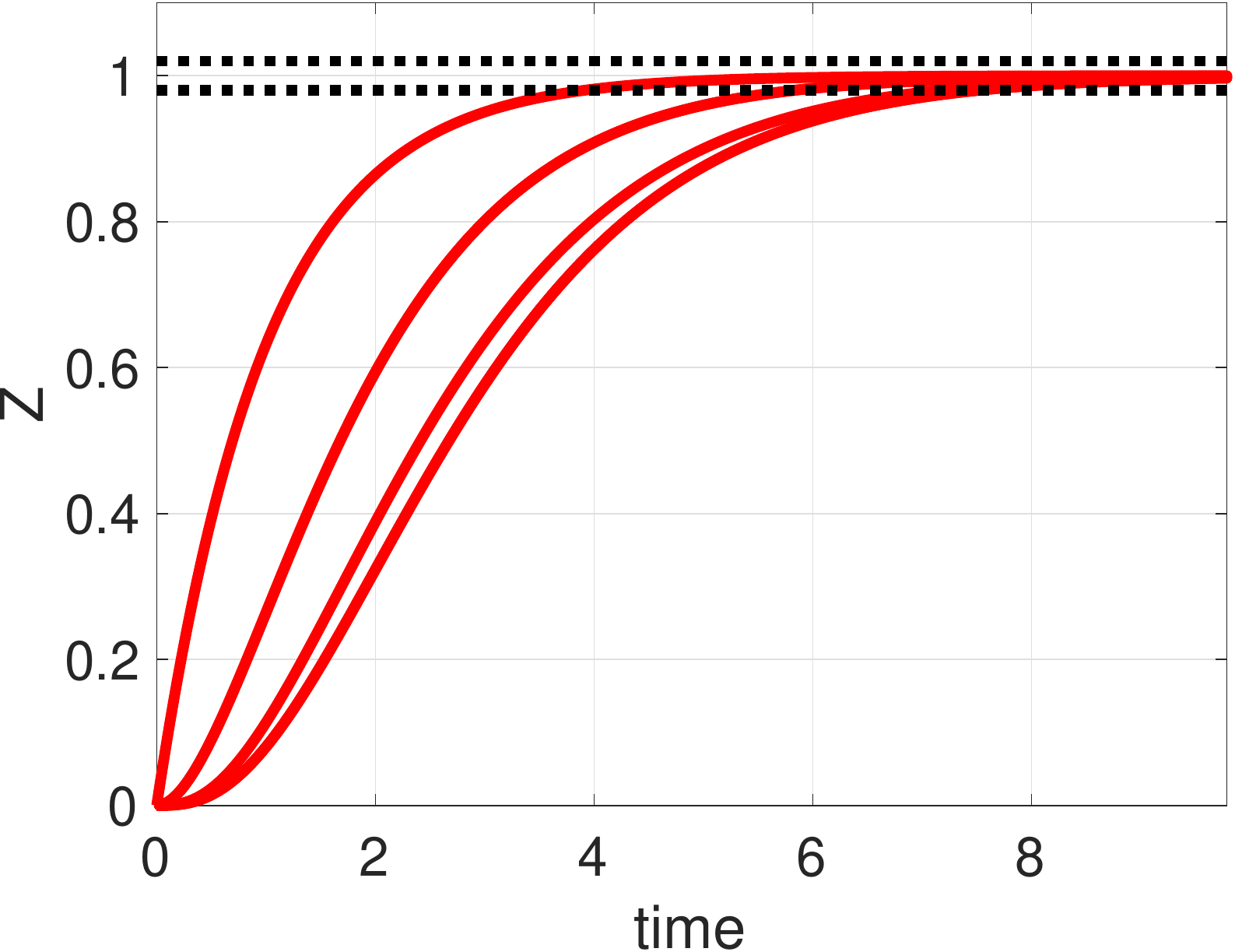}  & 
        \includegraphics[width=0.35\columnwidth]{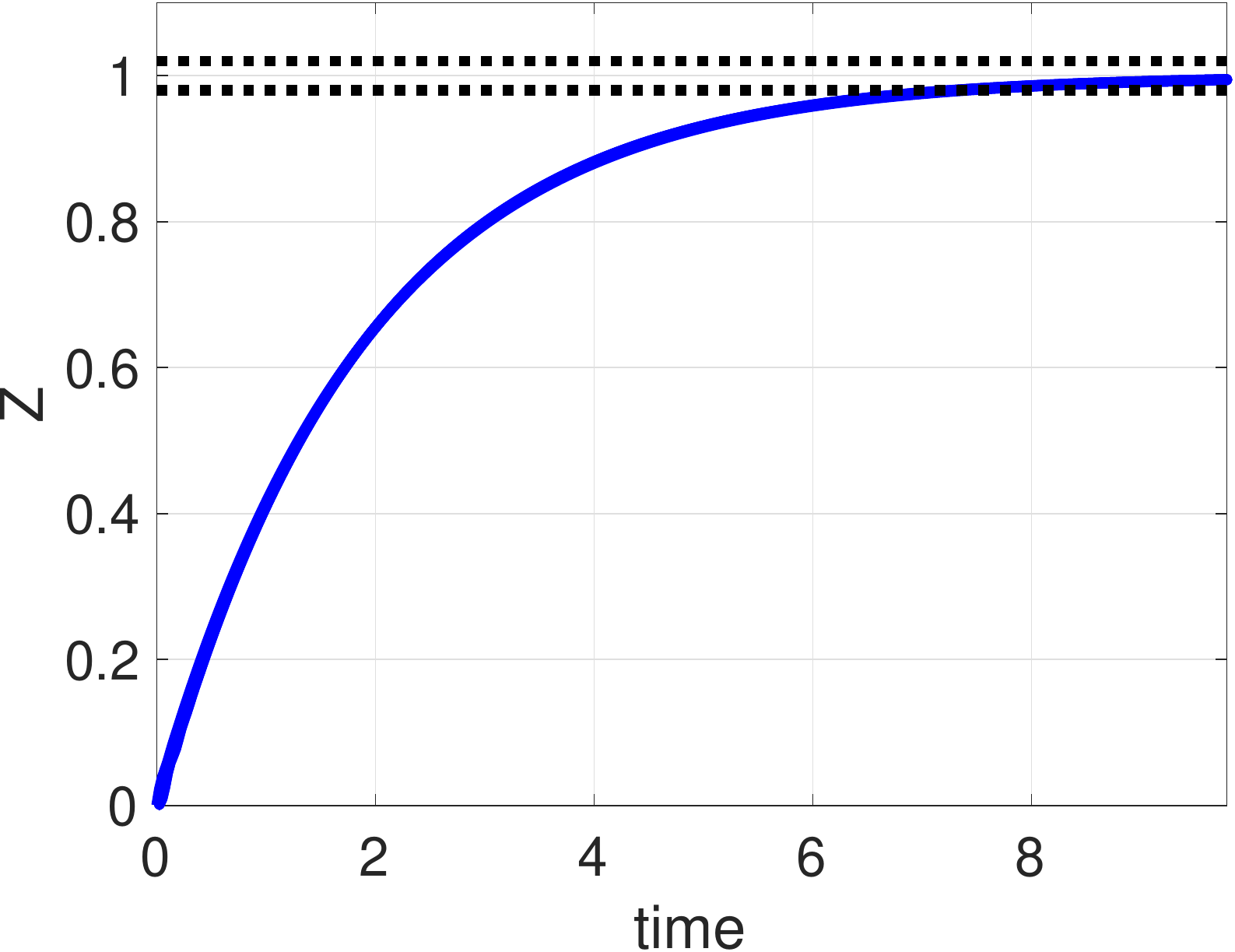}  
  \end{tabular}
\vspace{0.01in}
\caption{Step response $Z$ with first-order agents. (Left) Deviation from cohesion in step response of the nominal system without DSR in Eq.~\eqref{system_non_source}. 
(Right) Cohesive response with DSR as in Eq.~\eqref{system_DR}, with similar settling time $T_s$. }
\label{fig_results_fig2_vel_nodsr}
\end{center}
\end{figure}

  \vspace*{-0.1in}\noindent
To illustrate the  cohesion problem, the step response of the nominal system without DSR in Eq.~\eqref{system_non_source} with zero as initial condition  is shown in Fig.~\ref{fig_results_fig2_vel_nodsr}
where the the source $z_s$ is a unit step. 
Due to symmetry, and same initial conditions, states 2 and 3 are similar, and so are states 4 and 5, and hence there are four distinct plots in the 
step response. The loss of cohesion is visually observable in Fig.~\ref{fig_results_fig2_vel_nodsr} as differences between the different agent responses, and can be 
quantified as deviation $\Delta = 3.73$  in Eq.~\eqref{Eq_hl_cohesion}. The settling time $T_s$ to 2\% of the final value is  $T_s =  7.5$ s.
The normalized deviation  is $\Delta^* = 0.496$, as in  Eq.~\eqref{Eq_norm_hl_cohesion}. 

\suppressfloats

\subsection{Improved cohesion with DSR} 
\begin{figure}[!ht]
\begin{center}
\includegraphics[width=.35\columnwidth]{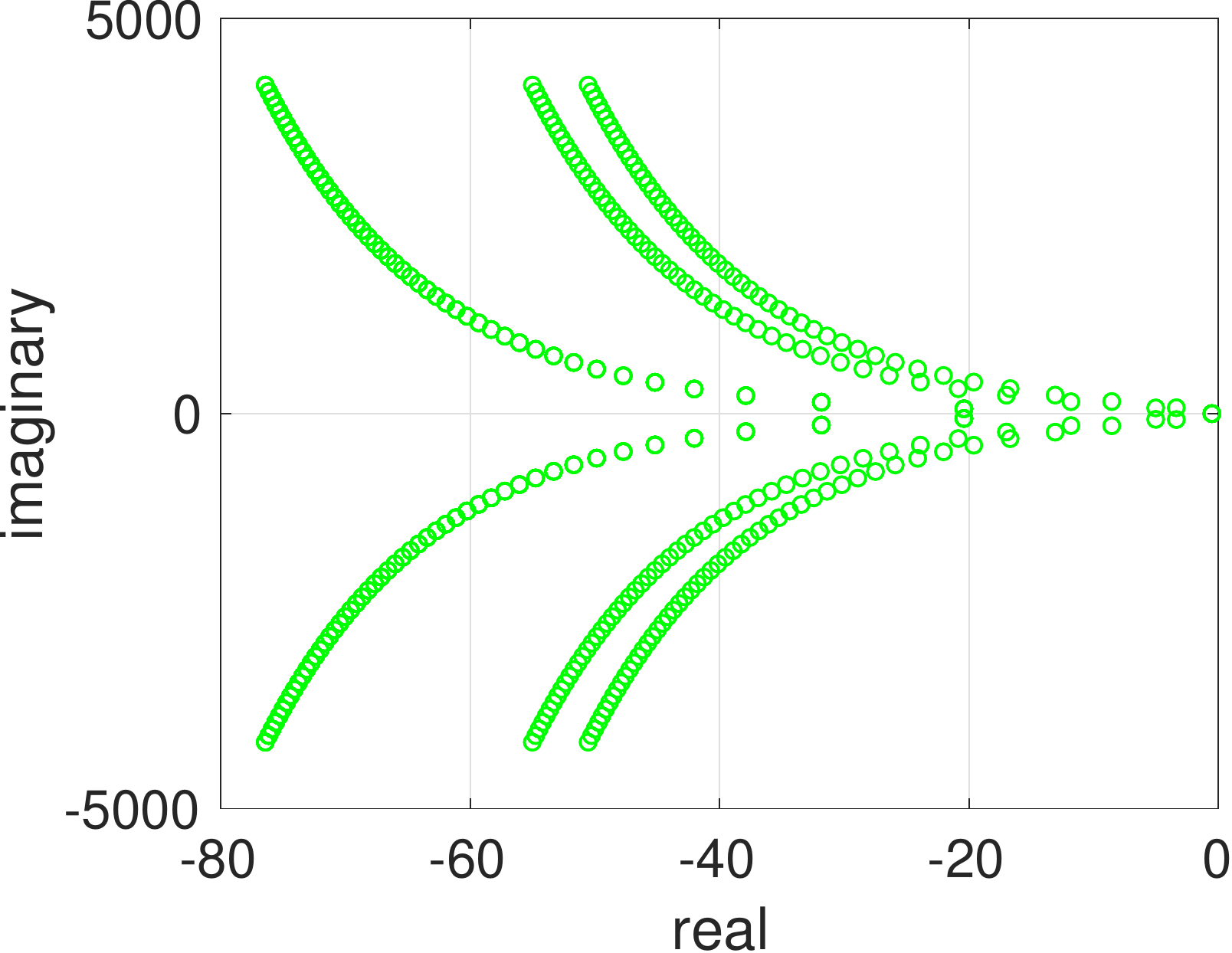}
\vspace{0.01in}
\caption{DDE in Eq.~\eqref{system_DR} is stable since the 
roots of its characteristic equation, found from  Eq.~\eqref{system_DR}, 
are in the open left-half of the complex plane.}
\label{fig1_poles_loc}
\end{center}
\end{figure}

  \vspace*{-0.1in}\noindent
The improvement in cohesion with DSR is evaluated when the response speed is similar to the case without DSR. Hence the parameter  $\alpha$  is selected to 
yield a similar settling time  $T_s =  7.5$ s as in Fig.~\ref{fig_results_fig2_vel_nodsr} for the case without DSR, i.e., from Eq.~\eqref{Settling_time}, $\alpha = 0.53$. 
Moreover, the delay $\tau$ is chosen to be a hundred times smaller than the settling time $T_s$ as in Remark~\ref{rem_Delayed_derivative_acts_filter}, i.e., 
$\tau = T_s/100 = 0.075$ s.
Since the eigenvalues of the pinned Laplacian $K$ are real, the DSR-based approach is stable if the parameter $\beta$ is chosen to be larger than the inverse of smallest eigenvalue magnitude (which is one
 for this example), i.e., to satisfy Eq.~\eqref{Condition_Assumption_stability_filter_4} 
 in Assumption~\ref{assumption_controller_parameter}, 
  the parameter $\beta$ needs to be larger than one and is selected as $\beta = 2$. 
Then, with these choices of parameters ($\alpha, \beta, \tau$), the roots $s_{i,\hat{k}}$ of the characteristic equation associated with the DDE in Eq.~\eqref{system_DR}, found using the   Lambert W function as in Eq.~\eqref{Lambert_W_function_roots}, 
are all in the left half of the complex plane as seen in Fig.~\ref{fig1_poles_loc}, and thereby, confirming the expected stability of the DDE.

\suppressfloats

Substantial improvement in cohesion of the step response of the system with DSR in Eq.~\eqref{system_DR}  is seen in Fig.~\ref{fig_results_fig2_vel_nodsr}.  
The cohesion  deviation $\Delta$   in Eq.~\eqref{Eq_hl_cohesion}  has reduced by $77.5$ times from $\Delta = 3.72$ without DSR to   $\Delta = 0.048$  with DSR. 
The settling time  to 2\% of the final value is similar, i.e.,  $T_s =  7.4$ s, and the normalized deviation $\Delta^*$  in  Eq.~\eqref{Eq_norm_hl_cohesion}
also reduces substantially (by 76 times) to  $\Delta^* = 0.0065$. Thus, the use of DSR results in substantial improvements in the cohesion.

\vspace{-0.1in}\subsection{Cohesive versus rapid synchronization} 
\label{subsection_cohesion_vs_rapid}
\begin{figure}[!ht]
\begin{center}
    \begin{tabular}{@{}cc@{}}
    \includegraphics[width=0.35\columnwidth]{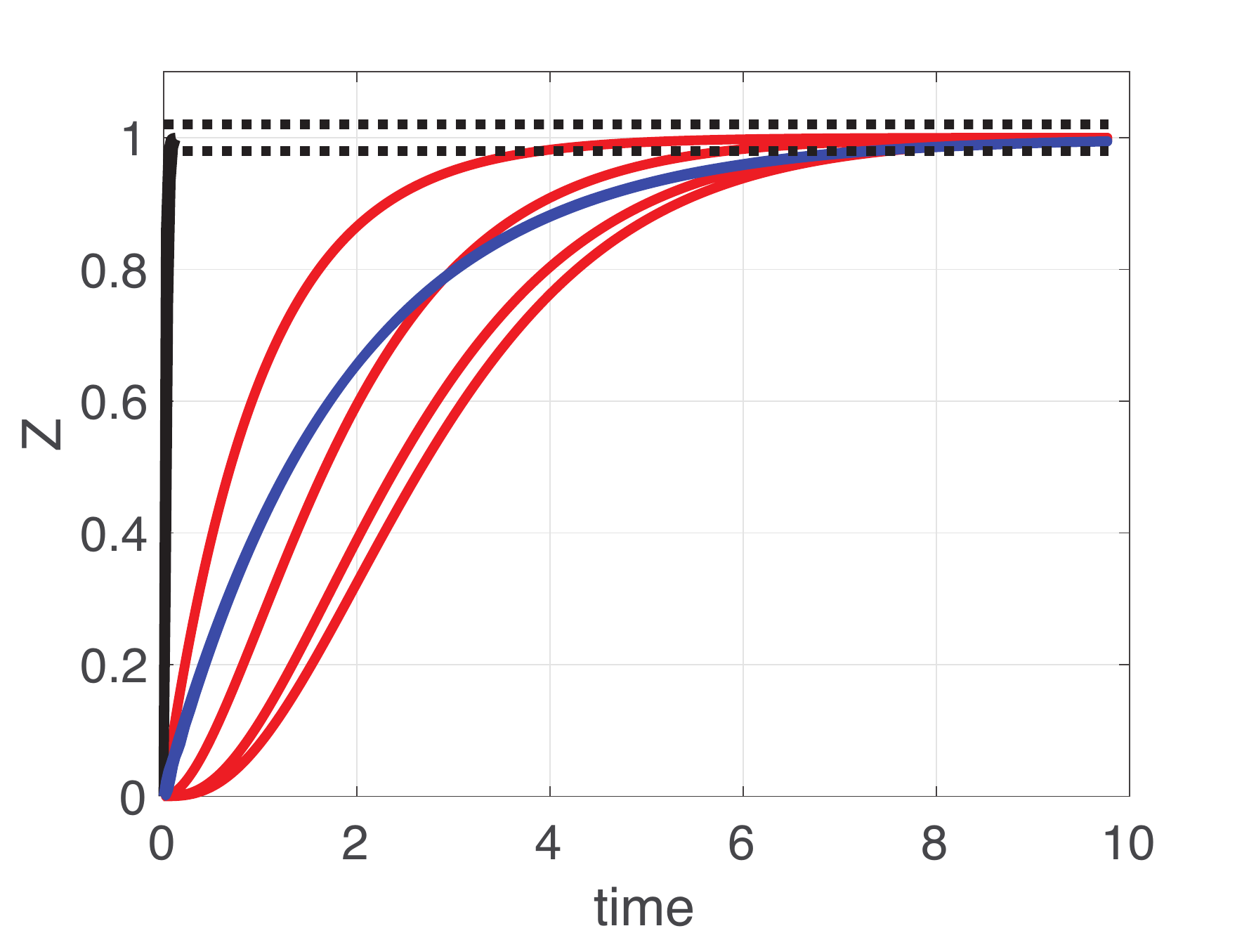}  & 
        \includegraphics[width=0.35\columnwidth]{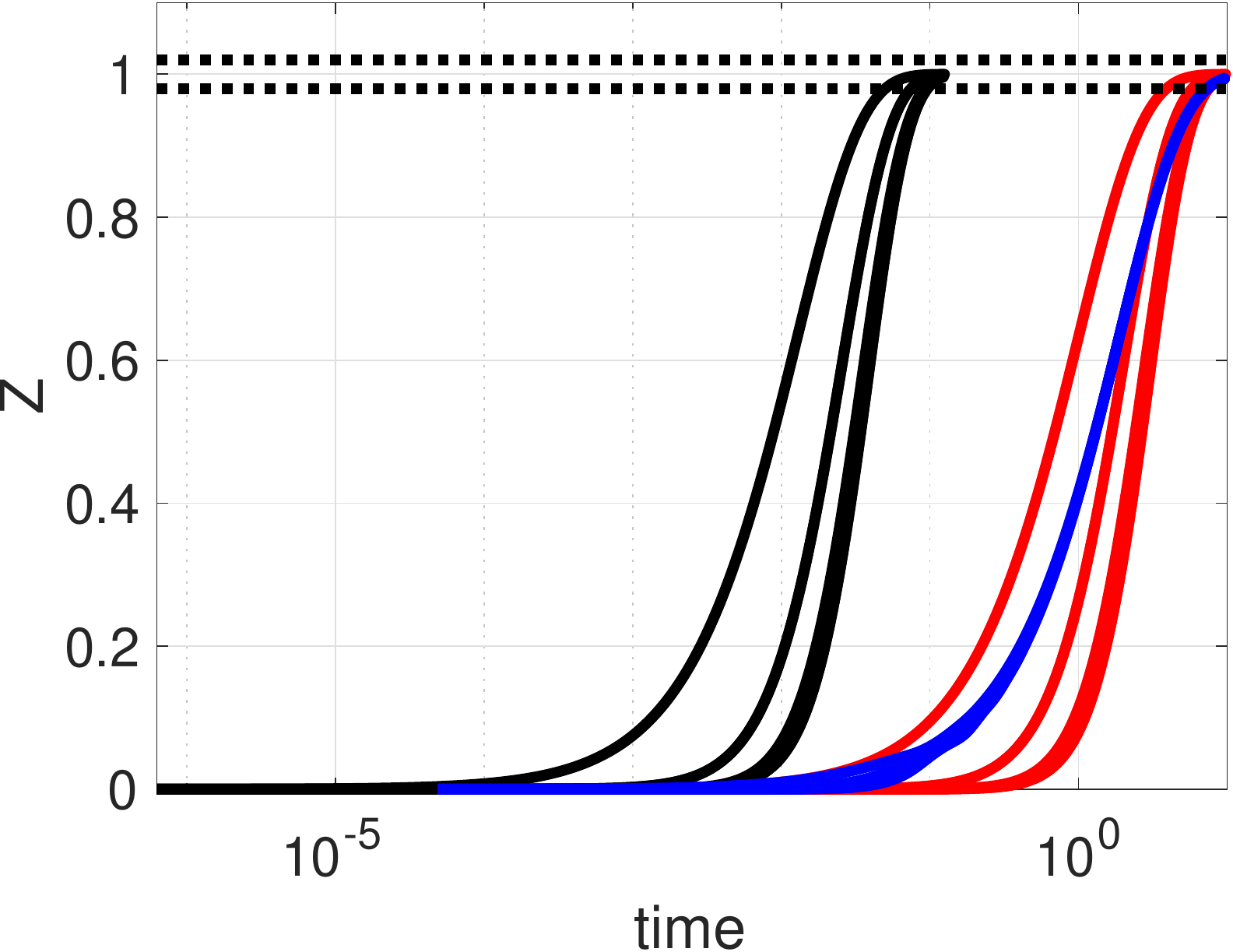}  
  \end{tabular}
\vspace{0.01in}
\caption{ Cohesive versus rapid synchronization of response $Z$ with 
first-order agents. Step responses illustrating 
cohesion improvement during the transition with DSR. 
Blue: with DSR. Red: with nominal $K$ and $B$ without DSR.  
Solid black: with scaled dynamics  $K$ and $B$ in Eq.~\eqref{system_non_source} and without DSR. 
(left) Faster response leads to a reduction of cohesion deviation $\Delta$ as seen by comparing the case with scaled dynamics (solid black lines) and 
without scaled dynamics (red lines). 
(right) The normalized deviation  $\Delta^*$  in  Eq.~\eqref{Eq_norm_hl_cohesion} 
is not reduced substantially by increasing the response speed 
by scaling up the dynamics, i.e., through rapid synchronization, 
as seen when the time is represented in a log scale. }
\label{fig_results_fig2_scaled}
\end{center}
\end{figure}

  \vspace*{-0.1in}\noindent
Without using the DSR approach, rapid synchronization can be used to reduce 
the difference between the responses by scaling the pinned Laplacian $K$ and the input matrix $B$ in 
Eq.~\eqref{system_non_source}  by the same factor $K_{gain}$ to speed up the convergence, i.e., 
\begin{equation}
\dot{Z} (t)  =  U    = -K_{gain} K Z (t) +K_{gain} B  z_s (t). 
\label{model_kgain}
\end{equation}
To match the deviation $\Delta$ in the response with DSR, 
the factor $K_{gain}$ was varied and the resulting deviation was numerically evaluated. The factor selection of $K_{gain} = 77.9$ led to a 
deviation in cohesion $\Delta =0.048,$ which is same as   $\Delta = 0.048$   for the case with the use of DSR. Note that this leads to a much faster response 
with a settling time of $T_s = 0.0964$ s when compared the settling time of $7.5$ s without scaling-up the dynamics, i.e., $K_{gain}=1$, as seen in  Fig.~\ref{fig_results_fig2_scaled}, with the time plotted in log scale. 

\suppressfloats

Nevertheless,  rapid synchronization does not result in cohesion during the transition in terms of the normalized deviation  $\Delta^*$  in  Eq.~\eqref{Eq_norm_hl_cohesion}. The responses remain substantially different from each other, as seen in Fig.~\ref{fig_results_fig2_scaled}. 
The normalized deviation $\Delta^*$  with rapid synchronization, achieved with a larger gain of 
$K_{gain}=77.9$, is  $\Delta^* = 0.496$, which is the 
same as the normalized deviation of  $\Delta^* = 0.496$ for the nominal case without DSR . 
Thus, scaling up with larger gain $K_{gain}$ leads to a faster response (rapid synchronization), 
but it does not lead to  improvements in the normalized cohesion $\Delta^*$.

\begin{figure}[!ht]
\begin{center}
\includegraphics[width=.6\columnwidth]{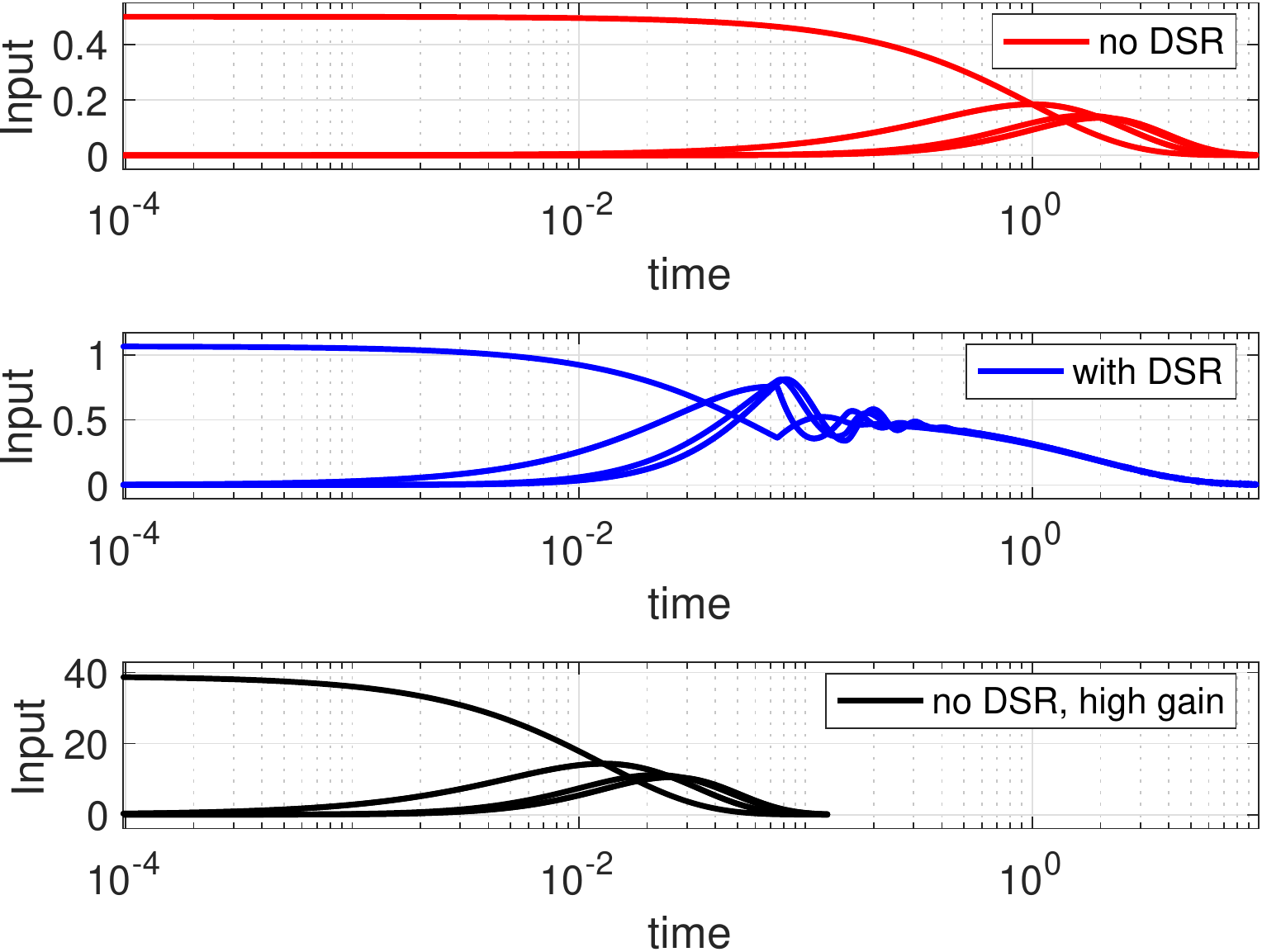}
\vspace{-0.1in}
\caption{Input $U$ with first-order agents. Maximum input without DSR (top) is $0.499$, with DSR the maximum input (middle) is about twice at $1.064$, and with 
the scaled-up dynamics in Eq.~\eqref{model_kgain} (bottom) is  $38.97$, which is $77.9$ times more that the nominal case without DSR (top).}
\label{fig3_input}
\end{center}
\end{figure}

Moreover, the rapid synchronization, achieved by scaling up the dynamics, requires a substantial increase in input magnitudes as seen in  Fig.~\ref{fig3_input}. 
The maximum input $U$ required without DSR is $0.499$, with DSR  is about twice at $1.064$, and with 
the scaled-up dynamics ($K_{gain}=77.9$) is  $38.97$, which is $77.96$ times more that the nominal case without DSR. 
Thus, the use of DSR increases the normalized cohesion without the substantial increase in input when compared to the rapid synchronization achieved by scaling-up of the dynamics.

\suppressfloats

\subsection{Example with higher-order dynamics} 
  \vspace*{-0.1in}\noindent
The impact of improved cohesion with DSR 
is  illustrated in the following for agents with higher-order dynamics.

\subsubsection{Higher-order dynamics example}
Consider a second order dynamics ($r=2$) for the agents, 
using the same network as in Fig.~\ref{fig_2_topological_ordering}, and therefore, the same pinned Laplacian $K$ and input matrix $B$ as in the first order example in Subsection~\eqref{subsection_example_system}. 
The system dynamics, with DSR,  
in Eq.~\eqref{system_DR_tracking_hod} becomes 
\begin{equation} 
\begin{aligned}
\ddot{Z} (t)   &  =   U(t)  = -\beta K \left [  2 \alpha \dot{Z}(t)  + \alpha^2 Z(t)  \right]  
 \\ &  \quad  \quad 
 + \beta    B  z_s^* (t)  ~
 + \left[ I - \beta K \right] \hat{{Z}}^{(2)}(t,\tau). 
\end{aligned}
\label{system_DR_tracking_hod_example}
\end{equation}
The DSR term $\hat{Z}^{(2)}(t,\tau)$  is as in Eq.~\eqref{system_DR_delayed_derivative_approx_hod}, 
\begin{align} 
\hat{Z}^{(2)}(s,\tau) & = \left[ f(s) \frac{ 1 - e^{-\tau s} }{\tau} \right]^2 Z(s) = {\mathcal{F}}(s) P(s), 
\label{system_DR_delayed_derivative_approx_hod_example}
\end{align}
with time domain representation, when the filter is selected as  $ f(s) =  \frac{\Omega}{s + \Omega}$, 
\begin{equation}
\begin{aligned} 
\frac{d}{dt} \hat{Z}^{(1)}(t,\tau) &  =  -\Omega  \hat{Z}^{(1)}(t,\tau)   + \Omega  \frac{ Z(t) - Z(t-\tau) }{\tau} \\
\frac{d}{dt}  \hat{Z}^{(2)}(t,\tau) &  =  -\Omega  \hat{Z}^{(2)}   + \Omega  \frac{ \hat{Z}^{(1)}(t) - \hat{Z}^{(1)}(t-\tau) }{\tau} .
\end{aligned}
\label{system_hod_example_2}
\end{equation}
The DSR parameters are selected similar to the  first-order case. 
The  delay is selected as 
$\tau = 0.075$ s and the parameter $\alpha= 1.195$. 
With the DSR-parameter $\beta=2$,  
the lower bound ${\epsilon_\lambda}$  is selected as in Eq.~\eqref{eq_hod_replace_overline_epsilon_lambda}, 
${\epsilon_\lambda}= {\overline{\epsilon_\lambda} } = 1.14$. 
The expression for stability in Eq.~\eqref{proof_rep_pole_5} was verified numerically as in Remark~\ref{rem_Computational_issues}, where 
the lhs  of Eq.~\eqref{proof_rep_pole_5}  was computed over the imaginary axis with the  
filter selected as a low pass filter as  
in Eq.~\eqref{system_hod_example_2}  and 
 $\Omega = \alpha/10$. 

\subsubsection{Results with higher-order dynamics}
\begin{figure}[!ht]
\begin{center}
    \begin{tabular}{@{}cc@{}}
    \includegraphics[width=0.35\columnwidth]{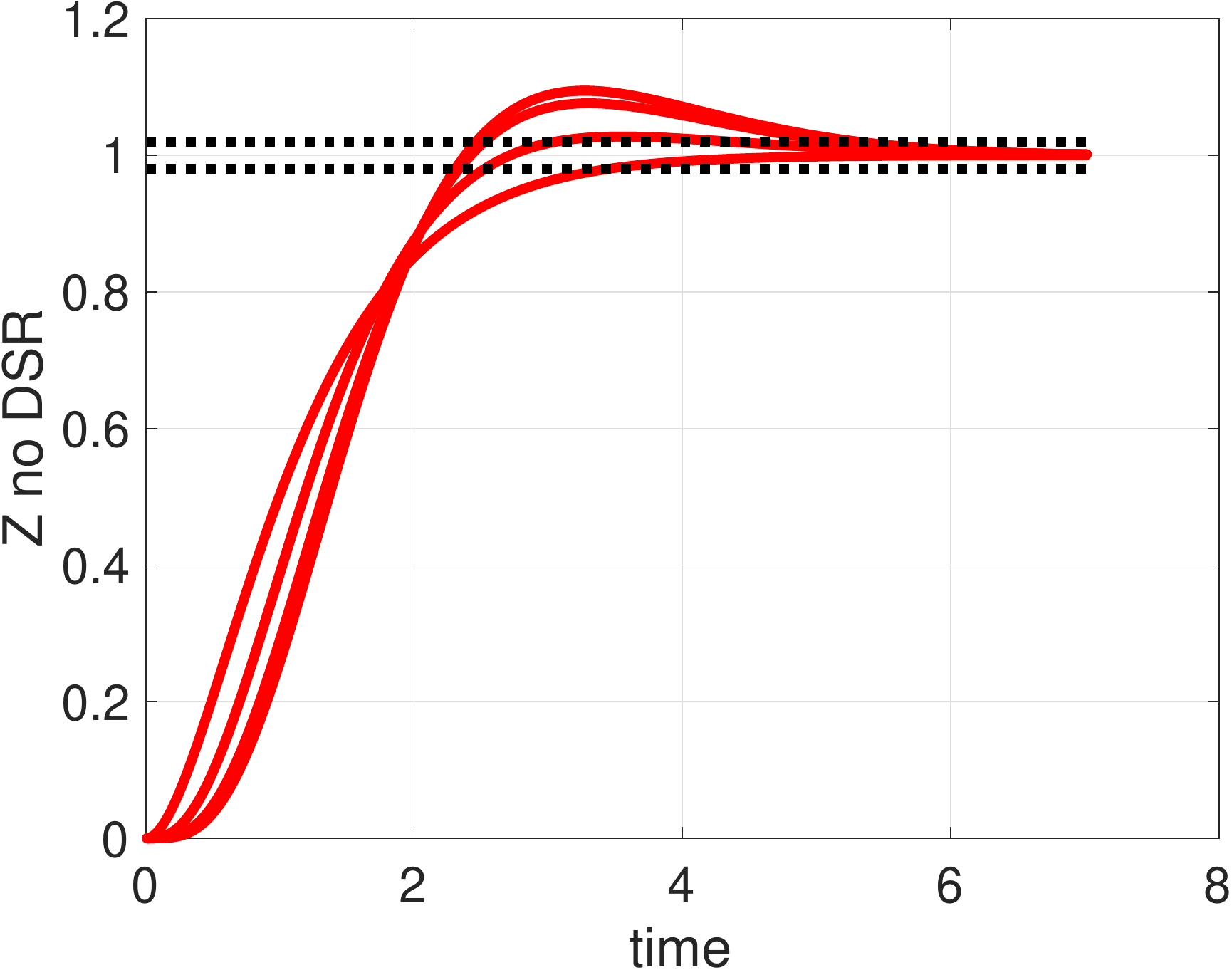}  & 
        \includegraphics[width=0.35\columnwidth]{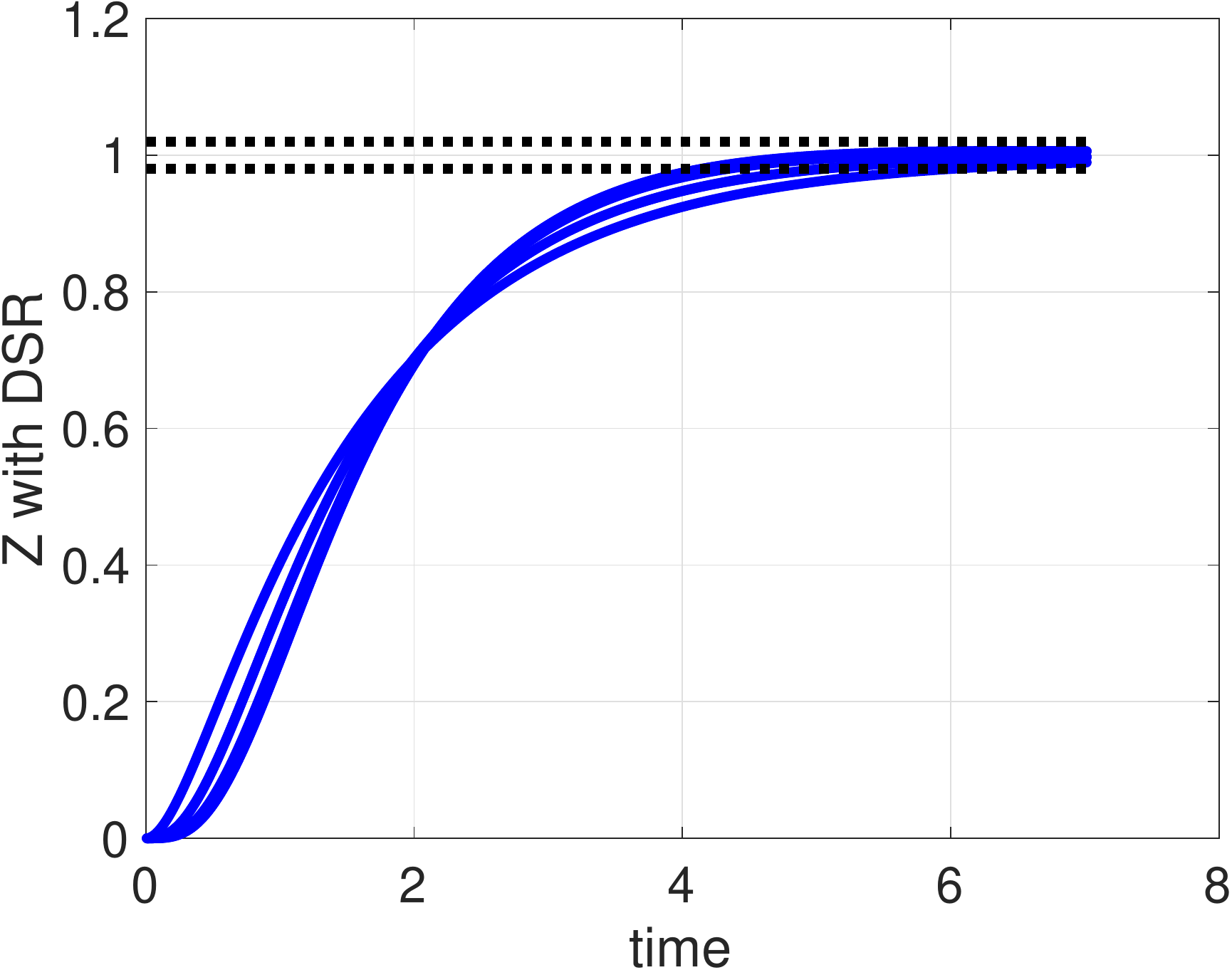}   \\
     \includegraphics[width=0.35\columnwidth]{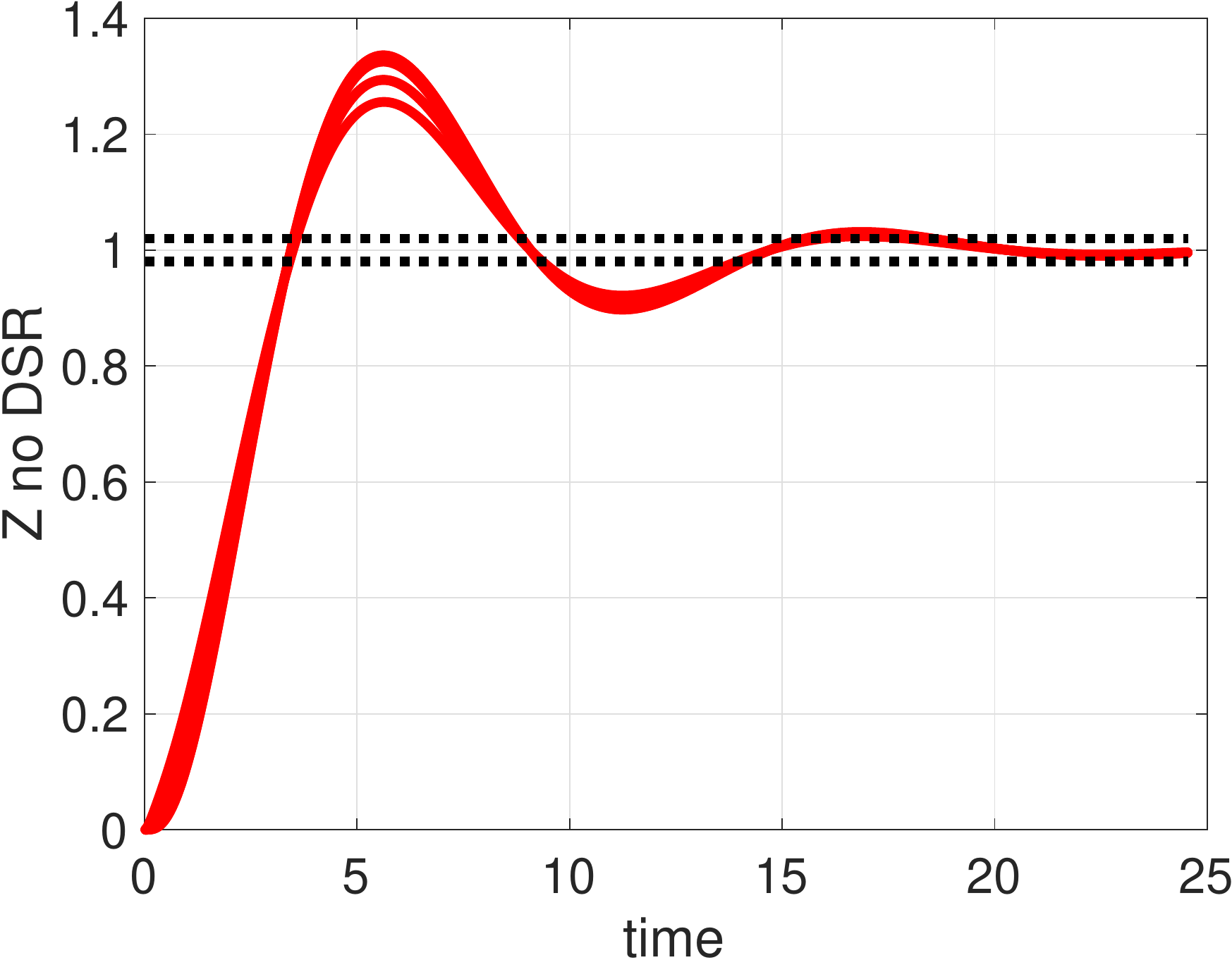}  & 
        \includegraphics[width=0.35\columnwidth]{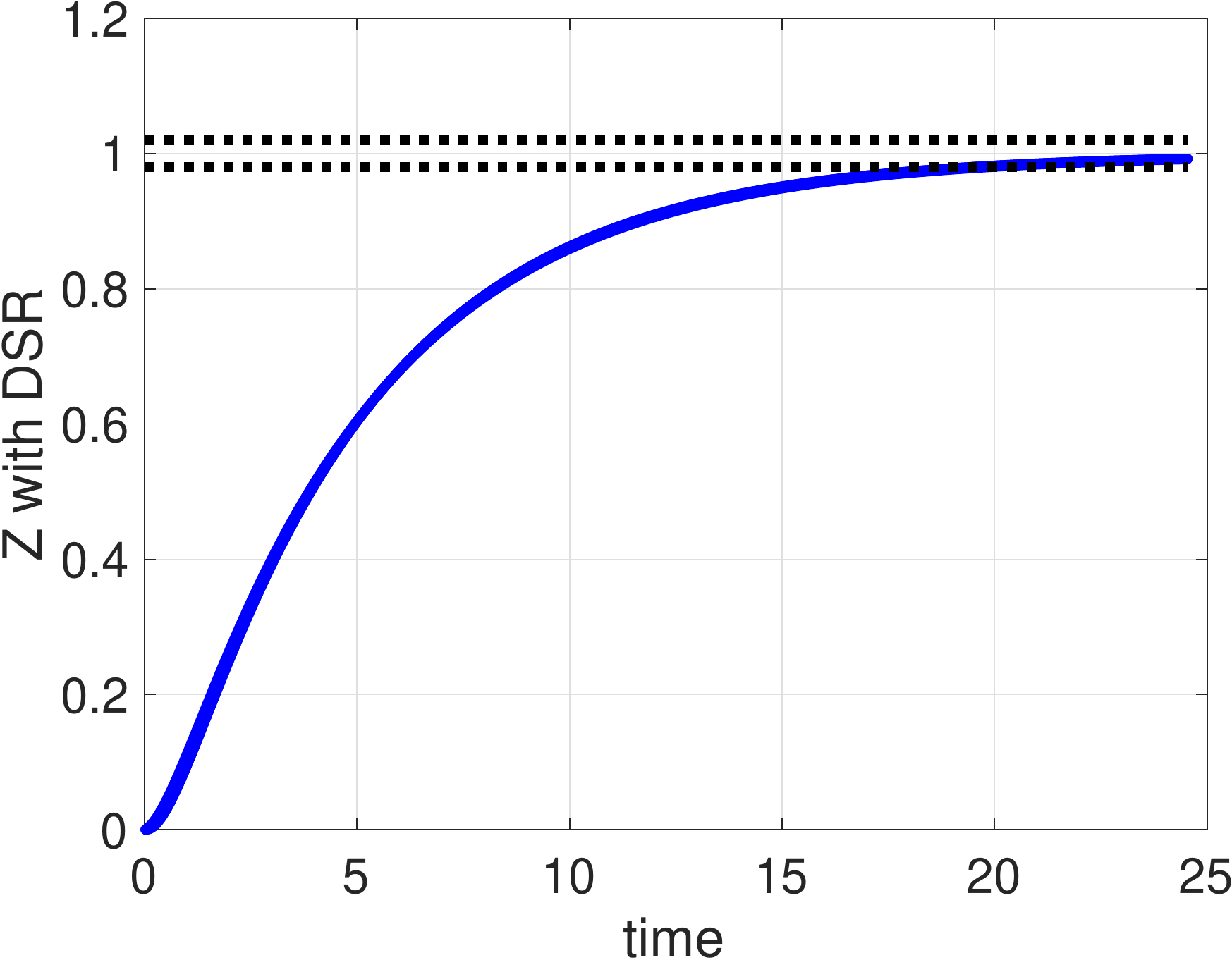}         
  \end{tabular}
\vspace{-0.01in}
\caption{Step response $Z$ with second-order agents
 with real eigenvalues (top) and complex eigenvalues (bottom). 
(Top left) Without DSR, i.e., 
$\hat{{Z}}^{(2)}(t,\tau) = 0$ and $\alpha = 1.69$  in Eq.~\eqref{system_DR_tracking_hod_example}. 
(Top right) More cohesive response with DSR as in Eq.~\eqref{system_DR_tracking_hod_example}.
(Bottom left) Without DSR, i.e., 
$\hat{{Z}}^{(2)}(t,\tau) = 0$ and $\alpha = 1.69$  in Eq.~\eqref{system_DR_tracking_hod_example}. 
(Bottom right) More cohesive response with DSR as in Eq.~\eqref{system_DR_tracking_hod_example}.
 }
\label{fig_results_hod_agents}
\end{center}
\end{figure}

\begin{figure}[!ht]
\begin{center}
    \begin{tabular}{@{}cc@{}}
    \includegraphics[width=0.35\columnwidth]{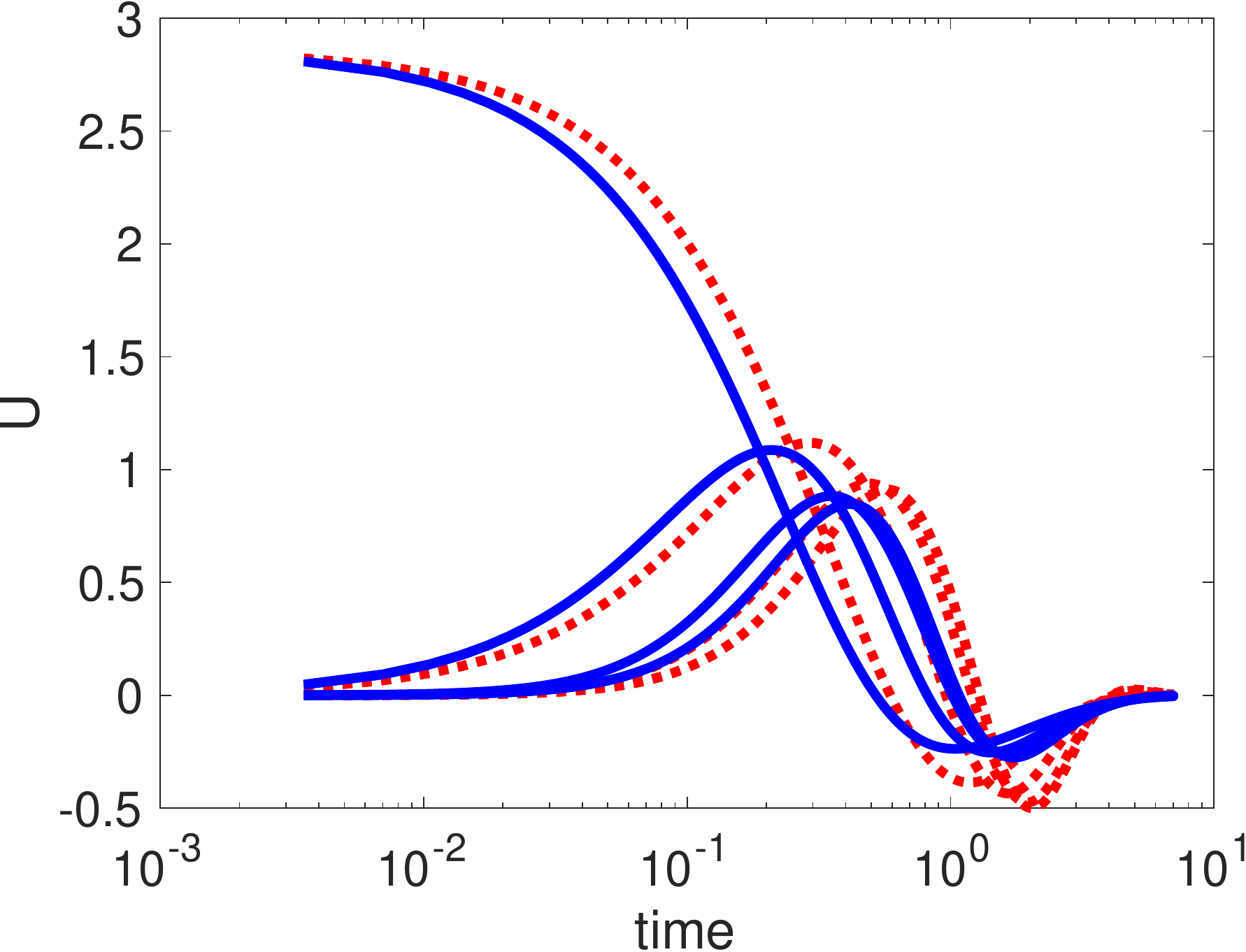}  & 
        \includegraphics[width=0.35\columnwidth]{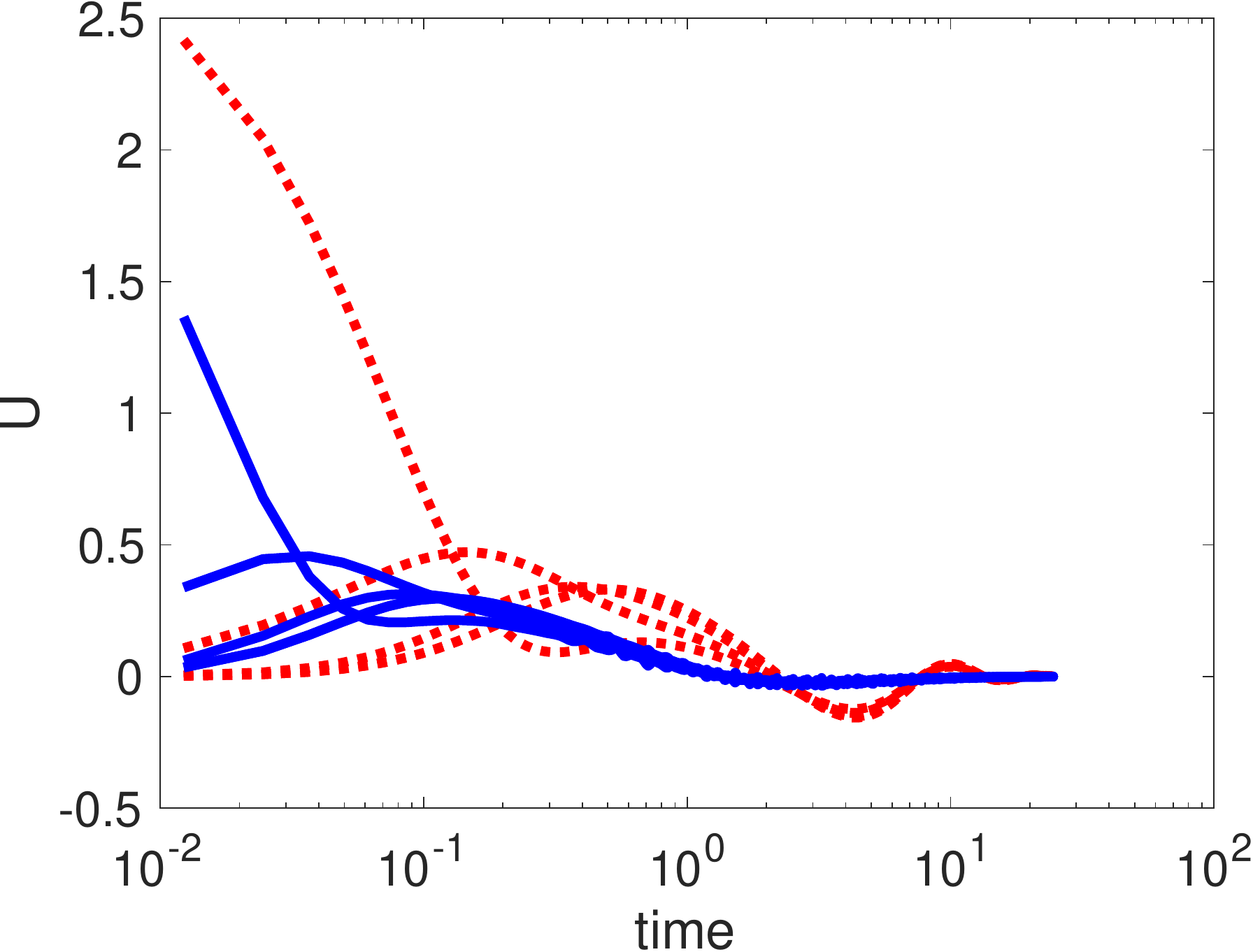}  
  \end{tabular}
 \vspace{-0.1in}
\caption{Input $U$ with second-order agents. (Left) real-eigenvalue case. 
Maximum input without DSR (red, dotted line) is $2.82$, which is similar to the case 
with DSR (blue, solid line) with a maximum input of $2.8$.
(Right) complex-eigenvalue case. 
Maximum input without DSR (red, dotted line) is $2.41$, and 
with DSR (blue, solid line) with a maximum input of $1.36$. }
\label{fig_input_hod_agents}
\end{center}
\end{figure}

The use of DSR leads to more cohesive response when compared to the case without DSR. 
The response with DSR is shown in Fig.~\ref{fig_results_hod_agents}. It is compared to the 
case without DSR, i.e., $\hat{{Z}}^{(2)}(t,\tau) = 0$  in Eq.~\eqref{system_DR_tracking_hod_example}
and a larger parameter $\alpha = 1.69$ so that the maximum input without DSR of $2.82$ is similar to the 
maximum input with DSR of $2.8$, as 
shown in Fig.~\ref{fig_input_hod_agents}.  The use of DSR results in similar settling time of $6$ s when compared 
to $5.3$ s without DSR. The cohesion deviation is reduced from $\Delta =1.04, ~\Delta^*=.17 $ without DSR to 
$\Delta =0.63, ~\Delta^*=0.11 $  with DSR. Thus, the simulations show that with similar-sized input, the DSR approach 
leads to increased cohesion for networked agents with higher-order dynamics.

\suppressfloats

\suppressfloats

\subsubsection{Results with complex eigenvalues}
Additional simulations were performed for higher-order dynamics (see Figs.~\ref{fig_results_hod_agents}, \ref{fig_input_hod_agents})  for the complex-eigenvalue case. 
This was achieved by adding edges from nodes $4,5,6$ to node $1$ 
in Fig.~\ref{fig_2_topological_ordering}. This leads to a loss of topological ordering, and leads to 
complex eigenvalues for the pinned Laplacian $K$. The resulting response tends to be more oscillatory.

Nevertheless,  reductions are seen in the cohesion deviation, 
as with the case for higher-order dynamics with real-valued eigenvalues. 
With 
the DSR-parameters $\tau = 0.075$ s,  $\beta=20$, and  $\alpha= 1.195$, 
the expression for stability in Eq.~\eqref{proof_rep_pole_5} was verified numerically as in Remark~\ref{rem_Computational_issues}. 
The use of DSR results in similar settling time of $19.9$ s when compared 
to $18.5$ s without DSR. The cohesion is improved from $\Delta =1.17, \Delta^*=0.058 $ without DSR to 
$\Delta =0.176, \Delta^*=0.009 $  with DSR. Thus, the simulations show that with similar settling time, the DSR approach 
leads to increased cohesion for networked agents with higher-order complex dynamics. 

The MATLAB code is attached as an  Appendix.

\subsection{Limitations and future work} 
  \vspace*{-0.1in}\noindent
Overall, the simulation results show substantial cohesion improvement with the  DSR approach, 
even for agents with higher-order dynamics.  
In general, network properties such as synchronization can be maintained 
if the network is jointly connected (instead of being always connected). 
Additional work  is needed to extend the current work to develop conditions for  
cohesion improvements  with the DSR approach for such time-varying networks. 
Similarly, there is potential to extend the DSR approach to improve 
cohesion during finite-time and fixed-time synchronization. 

\section{Conclusions}
\vspace*{-0.1in}\noindent
This work proposed  a new delayed-self reinforcement (DSR) approach  that enables cohesive response of multi-agent networks, during transitions from one cohesive operating point to another. 
Stability conditions were developed for the delay-based implementation of the proposed control law for general directed-graph networks. The potential for substantial improvements with the proposed DSR approach was illustrated with a simulation example and comparative analysis with and without DSR. 
The main advantage of the DSR approach is that it does not require reorganization of the network or increases in the response speed of the network to improve cohesion.  Moreover, the method is applicable to agents with higher-order heterogeneous dynamics.

\section{Acknowledgment}
\vspace*{-0.1in}\noindent
This work was partially supported by NSF grant CMII 1536306.

%

\clearpage

\newpage

\section*{Appendix}

{\small{

\begin{lstlisting}
% Simulations for problem in Automatica Paper
% Second order Complex Case system
% March 14 2020

    clear all 
    nfig=0;

%%%%%%%%%%%%%%%%%%%%%%%%%%%%%%%%%%%%%%%%%%%%%%%%%%%%%%%%%%%%%%%%%
    %%%%  Step 1: stimulate step response of the system without DSR
%%%%%%%%%%%%%%%%%%%%%%%%%%%%%%%%%%%%%%%%%%%%%%%%%%%%%%%%%%%%%%%%%

    N=6;  nleader=1;
    K = zeros(N,N);     % initialize the K matrix
    B = zeros(N,1);     % initialize the B matrix
        K(1,1:6)    = [3 0 0 -1 -1 -1];
        K(2,1:3)  = [-1  2 -1];
        K(3,1:3)  = [-1 -1  2];
        K(4,2:4)  = [-1 0  1]; 
        K(5,3:5)  = [-1 0  1];
        K(6,2:6)  = [-1  -1  -1  -1 4];
    % Pinning the leader 
    K(nleader,nleader) = K(nleader,nleader)+1;    
    B(nleader) = 1;
    
    % Need to choose beta to satisfy: \beta  > \frac{1}{\rho_{i} \cos{\phi_i} }, 
    % say 2 times the maximum of rhs over all i 
    lambda = eig(K);                    % the eigenvalues of K
    Ts = 7.5;
    tau   = Ts/100; 
    
    % similar settling time but for a second order system
    alpha_no_dsr = 1.69
    beta =20   
    alpha =alpha_no_dsr/sqrt(beta) 

    alpha_hat1 =  2*alpha; alpha_hat0 = alpha*alpha;
    alpha_hat1_nodsr =  2*alpha_no_dsr; 
    alpha_hat0_nodsr = alpha_no_dsr*alpha_no_dsr;
    
%%%%%%%%%%%%%%%%%%%%%%%%%%%%%%%%%%%%%%%%%%%%%%%%%%%%%%%%%%%%%%%%%
    %%%%  Step 2: simulate the system without the delayed control scheme Second Order System 
%%%%%%%%%%%%%%%%%%%%%%%%%%%%%%%%%%%%%%%%%%%%%%%%%%%%%%%%%%%%%%%%%
  
    % simulate the system without the delayed control scheme
    zK = zeros(size(K));  iK = eye(size(K)); % zero and identity matrices
    % define the system

    Aa = [zK iK; -alpha_hat0_nodsr*K -alpha_hat1_nodsr*K]; 
    Ba = [0*B; alpha_hat0_nodsr*B]; Ca = zeros(1,2*N); Ca(1,N)=1;  Da = []; 
    Cvel = zeros(1,2*N); Cvel(1,N)=1; 
    Sysnodelay = ss(Aa, Ba, Ca, Da);
    eigA = eig(Aa);

    vd= 1; % final desired state
    percentval_ts=2; % how to define final value 
    noiselevel = 0;

    % Find the setting time and redo with specify time vector
    [Y,t,X] = step(Sysnodelay); 
        tsteps = 2000;  % time steps in the simulations 
    tmax = max(t); t=1:1:tsteps; t = t*tmax/tsteps; 
    [Y,t,X] = step(Sysnodelay,t);
    X_no_dsr = X(:,1:N);
    V_no_dsr = X(:,N+1:2*N);
    U_no_dsr = -alpha_hat0_nodsr*K*X_no_dsr' -alpha_hat1_nodsr*K*V_no_dsr' +alpha_hat0_nodsr*B;

%%%%%%%%%%%%%%%%%%%%%%%%%%%%%%%%%%%%%%%%%%%%%%%%%%%%%%%%%%%%%%%%%
%%%%  Step 3: Stability of system with DSR
%%%%%%%%%%%%%%%%%%%%%%%%%%%%%%%%%%%%%%%%%%%%%%%%%%%%%%%%%%%%%%%%%

    eigK = eig(K);
    eps_lamdba = min(abs(beta*eigK./(1-beta*eigK)))
    %%% check for stability         
    Omega = .1*alpha;  % filter selection
    w  =1:Omega/10:100*Omega; 
    fs = Omega./abs((j*w) + Omega) ;
    lhs = (1-exp(-tau*j*w)).*fs;
    lhs = max(abs(lhs));
    rhs = tau*(eps_lamdba.^(1/2))*alpha;
    if rhs >= lhs
        lhs_rhs = [ lhs rhs]
        disp('stability condition is met')
    else
        lhs_rhs = [ lhs rhs]
        disp('Problem: stability condition is not met')
        return
    end
    

%%%%%%%%%%%%%%%%%%%%%%%%%%%%%%%%%%%%%%%%%%%%%%%%%%%%%%%%%%%%%%%%%
%%%%  Step 4: simulate the system with DSR
%%%%%%%%%%%%%%%%%%%%%%%%%%%%%%%%%%%%%%%%%%%%%%%%%%%%%%%%%%%%%%%%%

    Aadsr = [zK iK; -alpha_hat0*K -alpha_hat1*K]; 
    Badsr = [0*B; alpha_hat0*B]; Ca = zeros(1,2*N); Ca(1,N)=1;  Da = []; 
  
    Nsim = N*4;  dsr_or_not=1;  
    history = [zeros(Nsim,1)];
    Param_dde = [alpha tau beta N Omega dsr_or_not];
    %sol = dde23(ddefun,lags,history,tspan,options)
    sol = dde23(@ddefun,[tau],history,[0, tmax],[],Aadsr,Badsr,K,Param_dde);
    y_dsr = deval(sol,t);
    % Definitions in dde23 function dydt = ddefun(t,y,ydel,Aa,Ba,K,Param)
    %alpha = Param(1); tau = Param(2); beta = Param(3); 
    %N = Param(4); Omega = Param(5); dsr_or_not = Param(6); 
    
        
    X_dsr = y_dsr(1:N,:);
    V_dsr = y_dsr(N+1:2*N,:);
    Phat2_dsr = y_dsr(3*N+1:4*N,:);
    U_dsr = -beta*alpha_hat0*K*X_dsr -beta*alpha_hat1*K*V_dsr +beta*alpha_hat0*B  + (eye(size(K))-beta*K)*Phat2_dsr;

    
    nfig=nfig+1; figure(nfig)
    plot(t,X_no_dsr,'r:',t,X_dsr,'b','LineWidth',4);  
    xlabel('time'); ylabel('Z'); 
    set(gca,'FontSize',20)
    pause(0.01)
    % Blue(DSR) Red(no-DSR)
    %legend('no DSR','with DSR')

    nfig=nfig+1; figure(nfig)
    semilogx(t,U_no_dsr,'r:',t,U_dsr,'b', 'LineWidth',4); 
    xlabel('time'); ylabel('U');
    set(gca,'FontSize',20)
    pause(0.01)
    %Blue(DSR) Red(no-DSR)
    saveas(gcf,'fig_hdf_U_comparison_rev_resp','epsc')

    Max_input_U_dsr = max(max(U_dsr))
    Max_input_U_no_dsr = max(max(U_no_dsr))
    
    return
    
%%%%%%%%%%%%%%%%%%%%%%%%%%%%%%%%%%%%%%%%%%%%%%%%%%%%%%%%%%%%%%%%%
%   function to define the derivative for DDE solution
%%%%%%%%%%%%%%%%%%%%%%%%%%%%%%%%%%%%%%%%%%%%%%%%%%%%%%%%%%%%%%%%%

function dydt = ddefun(t,y,ydel,Aa,Ba,K,Param)
% Differential equation function
alpha = Param(1); tau = Param(2); beta = Param(3); 
N = Param(4); Omega = Param(5); dsr_or_not = Param(6); 

% define current and delayed states X=[P;Pdot], Phat1, Phat2
yd =1; 
X = y(1:N*2,1);

P = y(1:N,1);
Pdel = ydel(1:N,1);
Phat1 = y((N*2 +1):N*3,1);
Phat1del = ydel((N*2 +1):N*3,1);
Phat2 = y((N*3 +1):N*4,1);
Phat2del = ydel((N*3 +1):N*4,1);

dot_Phat1 = -Omega*Phat1 +Omega*(P -Pdel)/tau; 
dot_Phat2 = -Omega*Phat2 +Omega*(Phat1 -Phat1del)/tau; 
dot_X = [eye(N,N) zeros(N,N); zeros(N,N) beta*eye(N,N)]*Aa*X +beta*Ba*yd + dsr_or_not*[zeros(N,1); (eye(size(K))-beta*K)*Phat2];

dydt = [dot_X ; dot_Phat1; dot_Phat2];
end % ddex4de


   
\end{lstlisting}

}}

\end{document}